\documentclass[preprint,aps,showpacs,preprintnumbers,amsmath,amssymb,superscriptaddress,longbibliography]{revtex4-1}
\usepackage{graphicx}
\usepackage{dcolumn}
\usepackage{multirow}
\usepackage{placeins}
\usepackage{array} 
\usepackage{booktabs} 
\usepackage{xcolor}
\usepackage[title]{appendix}
\usepackage[normalem]{ulem}
\usepackage{float}
\usepackage[unicode=true, bookmarks=false, breaklinks=false, pdfborder={0 0 1},colorlinks=false]{hyperref}

\newcommand{\unit}[2][1]{#1~\mathrm{#2}}

\newcommand{\KTH}{Department of Applied Physics, School of Engineering Sciences, KTH Royal Institute of Technology, AlbaNova University Center, SE-10691 Stockholm, Sweden}
\newcommand{\WISEKTH}{Wallenberg Initiative Materials Science for Sustainability (WISE), KTH Royal Institute of Technology, SE-10044 Stockholm, Sweden}
\newcommand{\SeRC}{SeRC (Swedish e-Science Research Center), KTH Royal Institute of Technology, SE-10044 Stockholm, Sweden}

\begin{document}

\title{Mechanical control of magnetic exchange and response in GdRu$_2$Si$_2$: A computational study}


\author{Sagar Sarkar}\thanks{These authors contributed equally to this work}
\affiliation{Department of Physics and Astronomy, Uppsala University, Uppsala, 751 20, Sweden.}

\author{Rohit Pathak}\thanks{These authors contributed equally to this work}
\affiliation{Department of Physics and Astronomy, Uppsala University, Uppsala, 751 20, Sweden.}

\author{Arnob Mukherjee}
\affiliation{Department of Physics and Astronomy, Uppsala University, Uppsala, 751 20, Sweden.}

\author{Anna Delin}
    \affiliation{\KTH}
    \affiliation{\WISEKTH}
    \affiliation{\SeRC}

\author{Olle Eriksson}
\affiliation{Department of Physics and Astronomy, Uppsala University, Uppsala, 751 20, Sweden.}
\affiliation{Wallenberg Initiative Materials Science for Sustainability, Uppsala University, 75121 Uppsala, Sweden.}

\author{Vladislav Borisov}
\affiliation{Department of Physics and Astronomy, Uppsala University, Uppsala, 751 20, Sweden.}
\affiliation{Wallenberg Initiative Materials Science for Sustainability, Uppsala University, 75121 Uppsala, Sweden.}

\date{\today}

\begin{abstract}

We present a systematic computational study of the effect of uniaxial strain on the magnetic properties of GdRu$_2$Si$_2$, a centrosymmetric material known to host a field-induced skyrmion lattice (SkL). Using first-principles density functional theory, we first demonstrate the pronounced sensitivity of the exchange and anisotropy to specific structural distortions. These DFT-derived interactions are then integrated into a classical spin model to construct comprehensive magnetic phase diagrams under both compressive and tensile strain. Our key finding is that compressive strain ($\sim 2\%$) acts as an effective tuning parameter, substantially expanding the stability region of the $\vec Q_{100}$-driven topologically nontrivial phases. This results from the shifts in the critical magnetic fields and enhancement of the energy scale of the favored magnetic wave vector. In contrast, tensile strain induces a different magnetic ground-state by promoting a different magnetic ordering vector, $\vec Q_{110}$, leading to entirely distinct phase behavior. This work not only provides a quantitative understanding of the structural-magnetic coupling in GdRu$_2$Si$_2$ but also establishes strain engineering as a powerful approach to control and optimize topologically non-trivial magnetic phases in centrosymmetric magnets.

\end{abstract}

\maketitle

\section{Introduction}

Magnetic materials that host topological spin textures, such as skyrmions, have attracted immense interest due to their potential to revolutionize the future of next-generation spintronics \cite{wiesendanger2016nanoscale}, quantum computing \cite{psaroudaki2021skyrmion,psaroudaki2023skyrmion}, and high-density data storage. Typically, these textures stabilize in materials where structural inversion symmetry is broken. This broken symmetry often leads to the Dzyaloshinskii–Moriya interaction ($\text{DMI}$)~\cite{DZYALOSHINSKY1958241,Moriya_PhysRev_1960}, which then competes with Heisenberg exchange, giving rise to non-collinear spin structures like spirals and, eventually, skyrmions. While researchers have focused heavily on $\text{DMI}$-driven systems like $\text{MnSi}$ \cite{muhlbauer2009skyrmion} and $\text{Co}$-doped $\text{FeSi}$ \cite{munzer2010skyrmion}, there is growing interest in skyrmions observed in centrosymmetric materials. In these cases, stabilization comes from magnetic or geometric frustration, as seen in $\text{Co-Zn-Mn}$ compounds \cite{karube2018disordered,ukleev2021frustration} and $\text{Gd}_2\text{PdSi}_3$ \cite{kurumaji2019skyrmion}.

A particularly fascinating recent development is the observation of nanoscale ($\sim 2\,\text{nm}$) square skyrmion lattices ($\text{SkL}$) in GdRu$_2$Si$_2$~\cite{khanh2020nanometric, NDKhanh_AdvSci.9_2022}. This system is especially intriguing because it is a centrosymmetric material that lacks the typical geometric frustration and, crucially, lacks the structural elements needed for $\text{DMI}$. GdRu$_2$Si$_2$ has a layered structure where the magnetic Gadolinium ($\text{Gd}$) atoms are arranged in a square lattice, separated along the tetragonal $c$-axis by layers of $\text{Ru}$ and $\text{Si}$ atoms. The material is known to exhibit complex field-induced phase transitions that include spin spirals ($\text{SS}$) and the $\text{SkL}$ phase~\cite{khanh2020nanometric, NDKhanh_AdvSci.9_2022, GWood_PRB.107_2023}. The origin of non-collinearity in these spiral phases has been widely debated. Nomoto \textit{et al.} \cite{nomoto2020formation} initially suggested that inter-orbital frustration, namely a competition between the ferromagnetic interaction in the Gd $5d$ channel and the antiferromagnetic interaction in the Gd $4f$ channel, was the key, rather than the conventional Ruderman-Kittel-Kasuya-Yosida ($\text{RKKY}$) mechanism, which is well known in intrinsic and extrinsic metallic magnets~\cite{YKvashnin_PRL.116_2016, SSarkar_2Dmat.9_2022, SSarkar_PRB.110_2024}. Conversely, Bouaziz \textit{et al.} \cite{bouaziz2022fermi} used $\text{DFT}$ and atomistic spin dynamics ($\text{ASD}$) to propose that Fermi surface nesting generates quasi-two-dimensional $\text{RKKY}$ exchange interactions, which stabilize the $\text{SkL}$ in the presence of easy-axis anisotropy and a magnetic field. More recently, magnetic torque and resistivity measurements by Matsuyama \textit{et al.} \cite{matsuyama2023quantum} supported the $\text{RKKY}$ picture, confirming its relevance to the helical magnetism and $\text{SkL}$ phase through the de Haas–van Alphen ($\text{dHvA}$) and Shubnikov-de Haas ($\text{SdH}$) oscillations. 

Based on this context, our recent computational work \cite{SSarkar_PRB.112_2025} established that GdRu$_2$Si$_2$ is in reality a strong 3D magnet and the exchange frustration arises primarily from competition between interlayer (mainly $\text{FM}$) and intralayer (predominantly $\text{AFM}$) exchange interactions ($J_{ij}$). A minimal Hamiltonian incorporating these $J_{ij}$ and uniaxial anisotropy ($K_U$) provides an excellent qualitative description of the magnetic field-dependent phase transitions \cite{SSarkar_PRB.112_2025}. However, the exact topological nature of phases requires including weaker interactions, such as dipolar coupling for the zero-field properties \cite{SSarkar_PRB.112_2025} or biquadratic exchange for the $\text{SkL}$ (Phase II) or the Meron-like phase (Phase III), as demonstrated by Hayami \textit{et al.}~\cite{hayami2021square}. Although weak interactions are important, exchange ($J_{ij}$) and uniaxial anisotropy ($K_U$) remain the two essential parameters that govern the magnetic response and phase transitions in the system~\cite{ TNomoto_JAP.133_2023, bouaziz2022fermi, SSarkar_PRB.112_2025}. 

Following this fundamental understanding, our present study investigates the mechanical tunability of GdRu$_2$Si$_2$'s magnetism by controlling these microscopic interactions ($J_{ij}$, $K_U$) through uniaxial strain along the $c$-axis ($\epsilon_c$). We hypothesize that this approach offers an efficient way to modulate exchange frustration. For instance, compressive strain ($+\epsilon_c$) will alter the $\text{Gd-Gd}$ distances by decreasing inter-layer and increasing intra-layer separation. This directly modifies the corresponding $J_{ij}$ strengths, which in turn alter the critical balance that causes the exchange frustration. Tensile strain ($-\epsilon_c$) will induce the opposite effects. This mechanism allows for a continuous tuning of the magnetic properties and response, and is an efficient way to control the magnetic phases. Our motivation is supported by previous experimental work that reported enhanced pressure-driven stability of the $\text{SkL}$ phase in various systems, including the cubic ferrimagnet Cu$_2$OSeO$_3$~\cite{levatic2016dramatic}, MnSi~\cite{AChacon_PRL.115_2015}, and even Gd$_2$PdSi$_3$~\cite{SSpachmann_PRB.103_2021}, and very recently in GdRu$_2$Si$_2$ itself~\cite{LGries_PRB.111_2025}. Uniaxial strain is the most appropriate mechanical strategy here due to the layered structure of GdRu$_2$Si$_2$.

By performing extensive first-principles calculations combined with $\text{ASD}$ simulations, we provide a microscopic understanding of this phase stability by revealing the exact correlation between the microscopic parameters ($J_{ij}, K_U$) and the bulk magnetic response. Our results show that excessive uniaxial strain (both tensile and compressive) destabilizes the non-collinear phases, driving the system toward a uniform $\text{FM}$ state without exchange frustration, but with different characteristics. For all intermediate strains, exchange frustration leads to various non-collinear phases depending on the external magnetic field. Our results reveal that tensile strain induces a competition between the $\vec Q_{100}$ and $\vec Q_{110}$ wave vectors, resulting in a distinct set of magnetic ground states. Crucially, by explicitly simulating the magnetic response using the strain-modified parameters, we construct a detailed magnetic field ($B$) versus strain ($\epsilon$) phase diagram, mapping out a continuous range of mechanically stabilized magnetic phases. These findings establish a fundamental link between mechanical strain and the magnetic response in GdRu$_2$Si$_2$ and provide a roadmap for the strain-engineering of magnetic topological phases in centrosymmetric materials.

The rest of the paper is organized into three sections (II-IV). In Section II, details of our computational methodology are provided. In Section III, we present and discuss the results of the strain effect on the structural, electronic, and magnetic properties of GdRu$_2$Si$_2$. Finally, Section IV provides the concluding remarks.

\section{ Computational Methodology}
\label{metho}

Our methodology integrates multiple state-of-the-art computational tools. We first employed Density Functional Theory (DFT) codes, including Quantum Espresso (QE) and Vienna Ab initio Simulation Package (VASP), for structural optimization under strain and the calculation of electronic properties. Key magnetic parameters, such as exchange interactions ($J_{ij}$) and magnetocrystalline anisotropy ($K_U$), were determined using the Relativistic Spin-Polarized Toolkit (RSPt), which features a full-potential Linear Muffin-Tin Orbital (LMTO) implementation of DFT. These calculated parameters were then used as input for large-scale Atomistic Spin Dynamics (ASD) simulations, executed via the Uppsala Atomistic Spin Dynamics (UppASD) software, to map out the magnetic spin textures and phase diagrams. The following subsections provide a detailed description of the computational approach.

\subsection{QE and VASP for Structural and  Electronic properties}

We performed DFT calculations for electronic structure and structural optimization using the projected augmented wave (PAW) method implemented in QUANTUM ESPRESSO version~7.2, with PAW pseudopotentials generated from Pslibrary \cite{giannozzi2017advanced, dal2014pseudopotentials}. The generalized gradient approximation (GGA) in the Perdew-Burke-Ernzerhof (PBE) parameterization was used for the Exchange-Correlation (Ex-Corr) functional \cite{perdew_PhysRevLett.77.3865_1996, perdew_PhysRevLett.78.1396_1997}, unless otherwise stated. A kinetic energy cut-off of 80 Ry for the wave function and 800 Ry for the charge density was employed in the plane wave basis set. For reciprocal space integration, we used an automatically generated uniform grid of k-points with an offset of half a grid step. A $\Gamma$-centered Monkhorst-Pack~\cite{DJChadi_PRB.8_1973, JHMonkhorst_PRB.13_1976} k-mesh of $20\times20\times10$ ensured the convergence of the total energy and local moments. To take into account the localized nature of the Gd $4f$ states considered in the valence band, we used DFT+$U$ calculations \cite{anisimov_jpcm.9.48_1997, kotliar_RevModPhys.78.865_2006} in the rotationally invariant formulation of Liechtenstein \textit{et al.} \cite{liechtenstein_PhysRevB.52.R5467_1995}, where the Coulomb interaction parameters were set to $U = \unit[6.7]{eV}$ and $J_H = \unit[0.7]{eV}$ for the Gd $4f$ states, based on previous studies \cite{nomoto2020formation}. For geometry optimization, we fixed the $c$-axis at different strain levels ranging from $-6\%$ to $9\%$, with increments of 1\%, and allowed the lattice vectors within the $ab$-plane and the atomic positions to be optimized until the forces were less than 0.001 eV/\r{A} on each atom. 

Following this, the electron localization function (ELF)~\cite{BSilvi_Nat.371_1994} for QE-optimized systems was calculated using the Vienna Ab initio Simulation Package (VASP)~\cite{kresse_PhysRevB.47.558_1993,kresse_PhysRevB.49.14251_1994,kresse_PhysRevB.54.11169_1996,kresse_cms.6.15_1996}. Non-magnetic calculations were performed for these tasks, and the rest of the computational setting was kept similar to the QE calculations. 

\subsection{RSPt for Magnetic exchange and Anisotropy}

To determine the magnetic exchange interactions between the Gd moments ($J_{ij}$), we recalculated the electronic structure using the Full Potential Linear Muffin-Tin Orbital (FP-LMTO) method, implemented in the Relativistic Spin-Polarized Toolkit (RSPt) \cite{Wills1987, Wills2000, wills2010full, RSPt}. Similar to the QUANTUM ESPRESSO calculations, we employed the GGA-PBE exchange-correlation functional. The basis set included both valence and semicore states, specifically constructed from the $6s$, $6p$, and $5d$ orbitals of Gd, $5s$, $5p$, and $4d$ orbitals of Ru, and the $3s$, $3p$, and $3d$ orbitals of Si. For computational efficiency, highly localized and non-interacting Gd $4f$ states were treated as core states with scalar relativistic corrections, eliminating the need for DFT+$U$ corrections. The kinetic tail energies were set to $-0.1$, $-2.3$, and $1.5$ Ry. For Brillouin zone integration, we utilized a $\Gamma$-centered Monkhorst-Pack grid with a resolution of $32\times 32\times 16$ k-points. To calculate interatomic magnetic exchange interactions through the magnetic force theorem (MFT) \cite{LIECHTENSTEIN198765, Lichtenstein_PhysRevB_2000, Szilva2023}, we map the \textit{ ab initio} Kohn-Sham Hamiltonian (or DFT Hamiltonian) onto an effective classical Heisenberg Hamiltonian, expressed as:

\begin{equation} {H} = - \sum_{i\neq j} J_{ij}, \vec{e}_{i}\cdot\vec{e}_{j}. \label{eqn1} \end{equation}

In this equation, $(i,j)$ refers to the indices of the magnetic sites, $\vec{e}_i$ and $\vec{e}_j$ are unit vectors that indicate spin directions at the sites $i$ and $j$, respectively, and $J_{ij}$ represents the exchange interaction between these spins. Using RSPt's Green function-based method, we computed $J_{ij}$ values through the generalized non-relativistic expression:

\begin{equation} J_{ij}=\frac{T}{4} \sum_{n} \operatorname{Tr}\left[\hat{\Delta}_{i}\left(i \omega_{n}\right)\hat{G}_{ij}^{\uparrow}\left(i \omega_{n}\right) \hat{\Delta}_{j}\left(i \omega_{n}\right) \hat{G}_{ji}^{\downarrow}\left(i \omega_{n}\right)\right]. \label{eqn2} \end{equation}

Here, $T$ denotes the temperature, $\hat{\Delta}$ is the onsite exchange potential describing exchange splitting at sites $i$ and $j$, and $\hat{G}_{ij}^{\sigma}$ is the intersite Green’s function for spin $\sigma$ (either $\uparrow$ or $\downarrow$). The $n^\mathrm{th}$ fermionic Matsubara frequency is represented by $\omega_n$. These terms are matrix elements in both orbital and spin space, with the trace taken over orbital indices. The convergence of the exchange parameters was ensured by refining the $k$-mesh to $(52\times 52\times 26)$ for these calculations.

To calculate the magnetic anisotropy energy (MAE), we applied the force theorem~\cite{GDaalderop_PRB.41_1990, OLBacq_PRB.65_2002} with the following recipe. First, we performed the non-relativistic self-consistent density functional theory calculations as implemented in RSPt. The converged potential and charge density were then used for three sets of fully relativistic one-shot (non-self-consistent / single iteration) calculations where magnetic moments are aligned along the three mutually orthogonal $[100]$, $[010]$, and $[001]$ Cartesian directions. To calculate MAE, we took the difference between the eigenvalue sums of the in-plane ($[100]$, $[010]$) and out-of-plane ($[001]$) magnetic moment directions. The converged k-mesh of size $60\times60\times30$ was used for the MAE calculation.

\subsection{UppASD for Atomistic Spin Dynamics Simulations}

In the next step, spin textures were simulated by atomistic spin dynamics (ASD) using the Uppsala Atomistic Spin Dynamics (UppASD) package [\onlinecite{uppasd},\onlinecite{Eriksson2017}], where we solve the Landau-Lifshitz-Gilbert (LLG) equation [\onlinecite{Landau1935},\onlinecite{Gilbert2004}] for the atomic magnetic moments ($m_i$):
\begin{equation}
    \frac{d\Vec{m_i}}{dt} = - \frac{\gamma}{1+\alpha^2}\, \Vec{m}_i \times [\Vec{B}_i + \Vec{b}_i(t)] - \frac{\gamma}{m_i} \frac{\alpha}{1+\alpha^2}\, \Vec{m}_i \times (\Vec{m}_i \times [\Vec{B}_i + \Vec{b}_i(t)]).
     \label{eqn3}
\end{equation}

Here $i$ is again the index for magnetic sites. $\gamma$ is the gyromagnetic ratio, and $\Vec{b}_i(t)$ is a stochastic magnetic field with a Gaussian distribution. The magnitude of this field is related to the damping parameter $\alpha$, which helps bring the system into thermal equilibrium at temperature $T$. We use a time step of $\Delta t = \unit[0.1]{fs}$ for the annealing phase and $\Delta t = \unit[1]{fs}$ for the measurement phase in the UppASD calculations to solve these differential equations.

The effective field $\Vec{B}_i$ experienced by each spin at site $i$ is derived from the partial derivative of the Hamiltonian $H$ with respect to the local magnetic moment,
\begin{equation}
    \Vec{B}_i = - \frac{\partial H}{\partial \Vec{m}_i}.
     \label{eqn4}
\end{equation}
The Hamiltonian $H$ includes the following terms:
\begin{equation}
    H = - \frac{1}{2} \sum_{i \neq j} J_{ij}\, \vec{e}_i \cdot \vec{e}_j - K_U \sum_{i} \left( \vec{e}_i \cdot \vec{z}\, \right)^2 - \sum_{i} \vec{B}_\mathrm{ext} \cdot \vec{e}_i,
     \label{eqn5}
\end{equation}

where the first term describes the Heisenberg exchange interactions, with $i$ and $j$ being the site indices, and $J_{ij}$ the strength of the exchange interaction, obtained from our first principles calculations (Eqn.~\ref{eqn1}). The second term describes magnetic anisotropy, which in this case is the uniaxial anisotropy ($K_U$), and the last term corresponds to Zeeman splitting under an external magnetic field $\vec{B}_\mathrm{ext}$. For small $\vec{B}_\mathrm{ext}$, the most significant contribution to the Hamiltonian is typically the Heisenberg exchange interaction.

For our atomistic spin dynamics (ASD) simulations, we used a $N \times N \times N$ supercell, containing around $10^5$ total spins, and with periodic boundary conditions. As detailed in~\cite{SSarkar_PRB.112_2025}, the integer $N$ specifically depends on the properties of the exchange interactions and was therefore required to be varied between $30$ and $40$ depending on the strained system under consideration. This adaptive sizing was necessary to accommodate an integer number of spiral wavelengths inside the simulation box. This is crucial for accurately modeling complex magnetic textures, consistent with the approach established in~\cite{SSarkar_PRB.112_2025}. More details on the exact cell dimensions ($N$) used for each strained system are provided in Table~S1 of the Supplementary Material (SM)~\cite{sm_sarkar2025} (for example, $N=37$ results in $101,306$ total spins). We first performed simulated annealing to bring the system into thermal equilibrium, followed by an ASD measurement phase to obtain the spin texture after the system had evolved via the LLG equation and reached an energy minimum. Simulated annealing was performed at gradually decreasing temperatures of $\unit[200]{K}$, $\unit[100]{K}$, $\unit[50]{K}$, and $\unit[10]{K}$, with (20,000 - 26,000) spin dynamics sampling steps at each temperature. After these annealing steps, we performed 600,000 sampling steps during the measurement phase at $\unit[0]{K}$, so that the spin system can reach an equilibrium state at zero temperature and given external magnetic field.

\section{Results and Discussions}

Our results and discussions section is organized into three subsections. In Subsection A, the structural optimization scheme for simulating the uniaxial strain and its effect on the volumetric properties is discussed. In Subsection B, we present the strain-induced changes in the electronic properties of the system. Finally, in Subsection C, we present magnetic exchange and uniaxial anisotropy data as a function of strain, as well as data from ASD simulations showing the magnetic response ($M$ vs. $B$) as a function of strain. This data is then used to construct a strain-dependent $B$ vs $\epsilon_c$ magnetic phase diagram of GdRu$_2$Si$_2$, identifying various mechanically stabilized magnetic phases.

\subsection{Structural properties and effect of uniaxial strain}

To simulate uniaxial strain along the $c$ axis ($\epsilon_c$), similar to an experimental setup, we used a particular structural optimization scheme. This method optimizes the in-plane parameters ($a$ and $b$) for a series of fixed $c$ values. The experimental parameters could not be used directly as an unstrained reference, because even after full structural optimization (including lattice parameters, unit cell volume, and atomic positions), computational methods often do not perfectly reproduce the experimental values. Hence, to establish an unstrained structure in our study, we first performed a full structural optimization using different Exchange-Correlation (Ex-Corr) functionals. We then compared the optimized structural parameters, including lattice parameters, unit cell volume, and bond lengths, with experimental data. Spin-polarized calculations with ferromagnetic (FM) ordering of the Gd spins were considered for this purpose. Localized Gd $4f$ states were treated in the valence band within the DFT+U approach described in the methodology section. As shown in Table \ref{tab:str_table}, the GGA-PBE functional produced results that most closely matched the experimental values. Consequently, we selected the GGA-PBE+U method for all subsequent calculations, with the corresponding fully optimized structure serving as an unstrained reference for our system.

\begin{table}[ht!]
    \centering
    \renewcommand{\arraystretch}{1.2} 
    \setlength{\tabcolsep}{8pt} 
    \begin{tabular}{|c|c|c|c|c|c|}
        \hline
        \multirow{2}{*}{\textbf{Ex-Corr}} & \multicolumn{5}{|c|}{\textbf{Structural Parameters}} \\ \cline{2-6}
        & \textbf{a=b (\AA)} & \textbf{c (\AA)} & \textbf{Volume (\AA$^3$)} & \textbf{Si-Si (\AA)} & \textbf{Si-Ru (\AA)} \\ \hline
        \textit{GGA-PBE}      & \textit{4.16143} & \textit{9.58672} & \textit{166.018} & \textit{2.519} & \textit{2.371} \\ \hline
        PBE-Sol      & 4.11697 & 10.0054 & 169.586 & 2.656 & 2.369 \\ \hline
        PBE+DFTD3-0~\cite{grimme2010consistent}  & 4.12377 & 9.60724 & 163.376 & 2.491 & 2.364 \\ \hline
        \textbf{Expt.}~\cite{khanh2020nanometric}        & \textbf{4.16340} & \textbf{9.61020} & \textbf{166.582} & \textbf{2.403} & \textbf{2.403} \\ \hline
    \end{tabular}
    \caption{Comparison of lattice parameters (\(a\), \(b\), \(c\)), unit cell volume, Si-Si bond distance, and Si-Ru bond distance obtained after a full structural optimization using different Ex-Corr functionals.  The corresponding experimental values from Ref.~\cite{khanh2020nanometric} are also tabulated for comparison purposes.}
    \label{tab:str_table}
\end{table}

\begin{figure} [H]
    \includegraphics[scale=0.50]{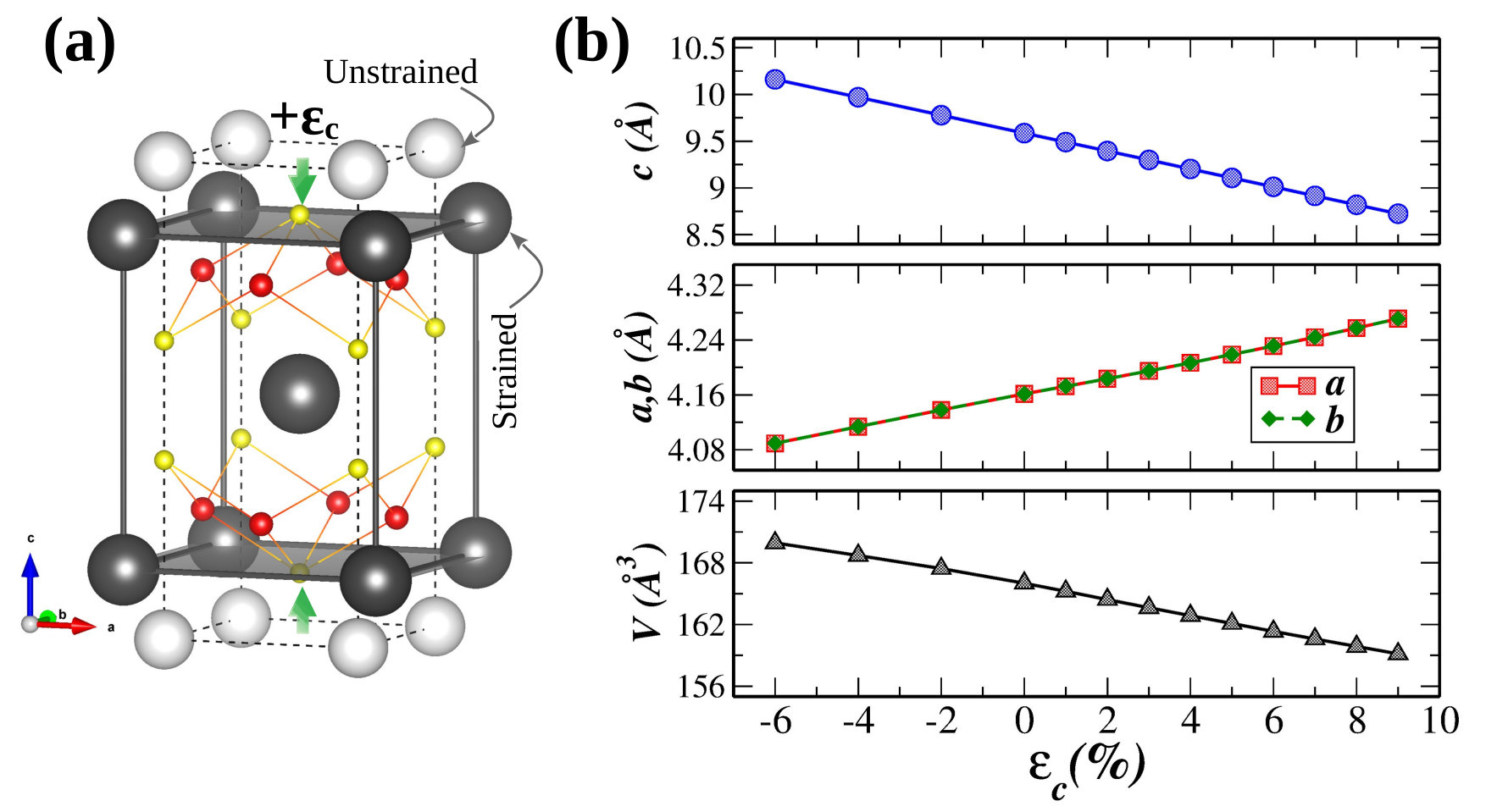}
    \caption{ (a) A schematic showing the effect of compressive uniaxial strain along $c$ axis ($+\epsilon_c$ indicated by the green arrows). The Gd-Gd interlayer distance decreases, whereas the intralayer Gd-Gd distances increase. The Gd atoms in the unstrained/strained state are shown with light grey/black spheres, respectively.  (b) Variation of unit cell lattice parameters (Top and Middle panels), and unit cell volume (Bottom panel) under uniaxial strain along $c$ axis ($\epsilon_c$). '+'/'-' values indicate compressive and tensile strains, respectively. The zero strain values are from our unstrained GGA-PBE optimized structure reported in Table~\ref{tab:str_table}. }
    \label{lattice-geo}
\end{figure}
\FloatBarrier

Next, the $c$ parameter of our unstrained structure was systematically compressed and expanded by various percentages in steps of $1\%$. For each strain level, we optimized the in-plane lattice parameters ($a$ and $b$) and atomic positions. By allowing $a$ and $b$ to relax freely, this simulation also becomes analogous to applying an in-plane biaxial strain. This is because the uniaxial strain along the $c$-axis induces a corresponding perpendicular strain in the $ab$-plane (known as the Poisson effect), as shown schematically in Fig.~\ref{lattice-geo} (a). For instance, an applied $c$-axis tensile strain is equivalent to an in-plane compressive biaxial strain, and the $c$-axis compressive strain is equivalent to an in-plane tensile biaxial strain. This equivalence is highly valuable because biaxial strain is often more natural and experimentally achievable, for example, by growing thin films on a lattice-mismatched substrate. This perspective supports the physical possibility of achieving the strain limits explored in our study. The resulting variation in the lattice parameters and unit cell volume as a function of uniaxial strain ($\epsilon_c$) is presented in Fig.~\ref{lattice-geo} (b). Compressive strain led to a decrease in $c$ and an increase in $a$ and $b$, accompanied by a reduction in unit cell volume, as expected.  Trends reverse in the case of tensile strain. The consistent equality and smaller variation of $a$ and $b$ compared to $c$ indicated preserved tetragonal crystal symmetry, which was confirmed by explicit symmetry checks using FINDSYM \cite{stokes2005findsym}. The absence of strain-induced phase transitions in Fig.~\ref{lattice-geo}(b) further confirmed that our applied strain range was within the elastic limit. The range of applied strain, from $-6\%$ tensile to $+9\%$ compressive, was specifically selected to investigate its profound impact on magnetic interactions. This range is particularly significant because $6\%$ tensile strain induces weak ferromagnetism, while the $9\%$ compression produces strong ferromagnetism with respect to changes in magnetic exchange. The intermediate systems within these limits exhibit varying degrees of exchange frustration, which will be discussed comprehensively in the magnetic properties section.

\subsection{Electronic properties and effect of uniaxial strain}

\begin{figure}[ht]
\includegraphics[scale=0.50]{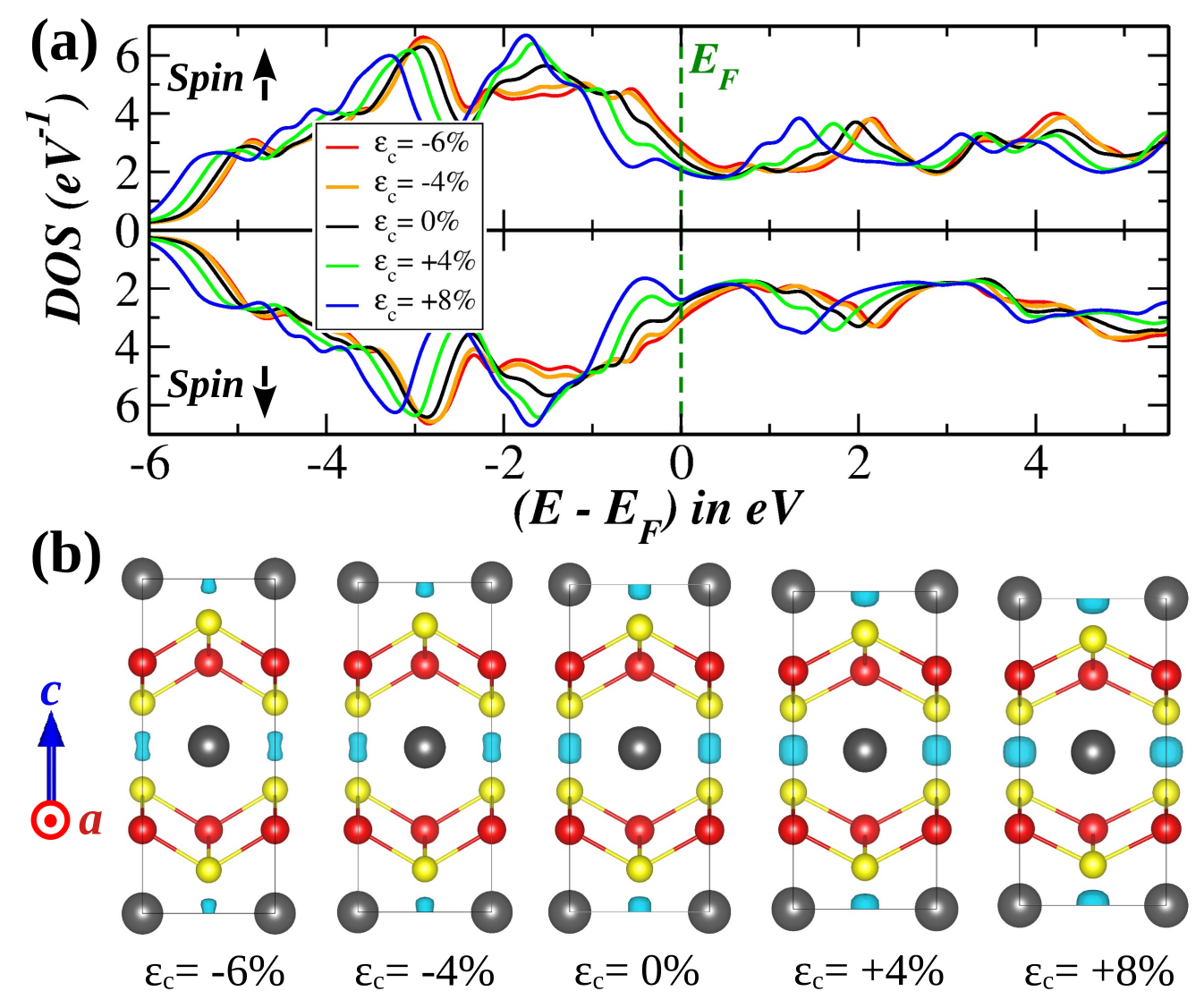}
\caption{(a) Calculated total DOS for different strained systems with FM spin order of the Gd moments. (b) ELF for an isosurface level of $\eta$ = 0.80 (in cyan colour), showing the location and qualitative amount of localized electrons in the unit cell for different strained systems. The dark grey spheres represent a Gd atom, while the smaller red and yellow spheres indicate the positions of
Ru and Si atoms, respectively, in the unit cell. An ELF with $\eta$ = 0.80 - 1.00 generally shows the localized electrons like lone-pairs, core-shell electrons, and covalent bond electrons. See text for details.}
\label{elf-fig}
\end{figure}
\FloatBarrier

Here, we explore how strain affects the electronic behavior of the system. We examine the total density of states (DOS) of our system in its unstrained state and under various tensile ($-6\%$ and $-4\%$) and compressive ($+4\%$ and $+8\%$) strains. We used the same computational method for these calculations as we did for optimizing the structure. Our findings, illustrated in Fig.~\ref{elf-fig} (a), show that the material remains metallic under all strained conditions. Interestingly, the contribution of electrons at the Fermi energy (a key indicator of metallic behavior) increases with tensile strain and decreases with compressive strain. This suggests that tensile strain enhances the metallic character, while compressive strain reduces it. To further understand how strain alters electronic bonding, we calculated the electron localization function (ELF)~\cite{BSilvi_Nat.371_1994} for these systems using VASP (see section II (a) for details), presented in Fig.~\ref{elf-fig} (b). These data were collected from non-magnetic calculations, but the computational settings were kept equivalent to the QE calculations. The ELF helps us to visualize the electron distribution. An ELF isosurface level ($\eta$) value between 0.80 and 1.00 typically indicates the presence of localized electrons, such as those involved in covalent bonds, lone pairs, or core shells\cite{BSilvi_Nat.371_1994, YGirin_Wiley.10_2014, HLevamaki_PRB.103_2021}. From the ELF data, we observe that tensile strain leads to weaker covalent bonds between silicon (Si) atoms, while compressive strain results in stronger covalent bonds. This aligns with our expectations, as tensile strain increases the Si-Si interlayer bond distance, and compressive strain decreases it. These localized covalent bonds are mainly formed by the electron states of Si $3s$ and $3p$. In contrast, the delocalized free electrons responsible for the metallic behavior (Fig.~\ref{elf-fig} (a)) primarily originate from the Ru $4d$ and Gd $5d$ electron states. Taken together, the electronic data in Fig.~\ref{elf-fig} indicate that tensile strain transforms the system into a weakly bonded layered metal, while compressive strain promotes a strongly bonded three-dimensional bulk metal.

\subsection{Magnetic properties and effect of uniaxial strain}

In this section, we present our main results showing the effect of strain on exchange frustration, anisotropy, and magnetic response of the system. 

\begin{figure}[ht]
\includegraphics[scale=0.5]{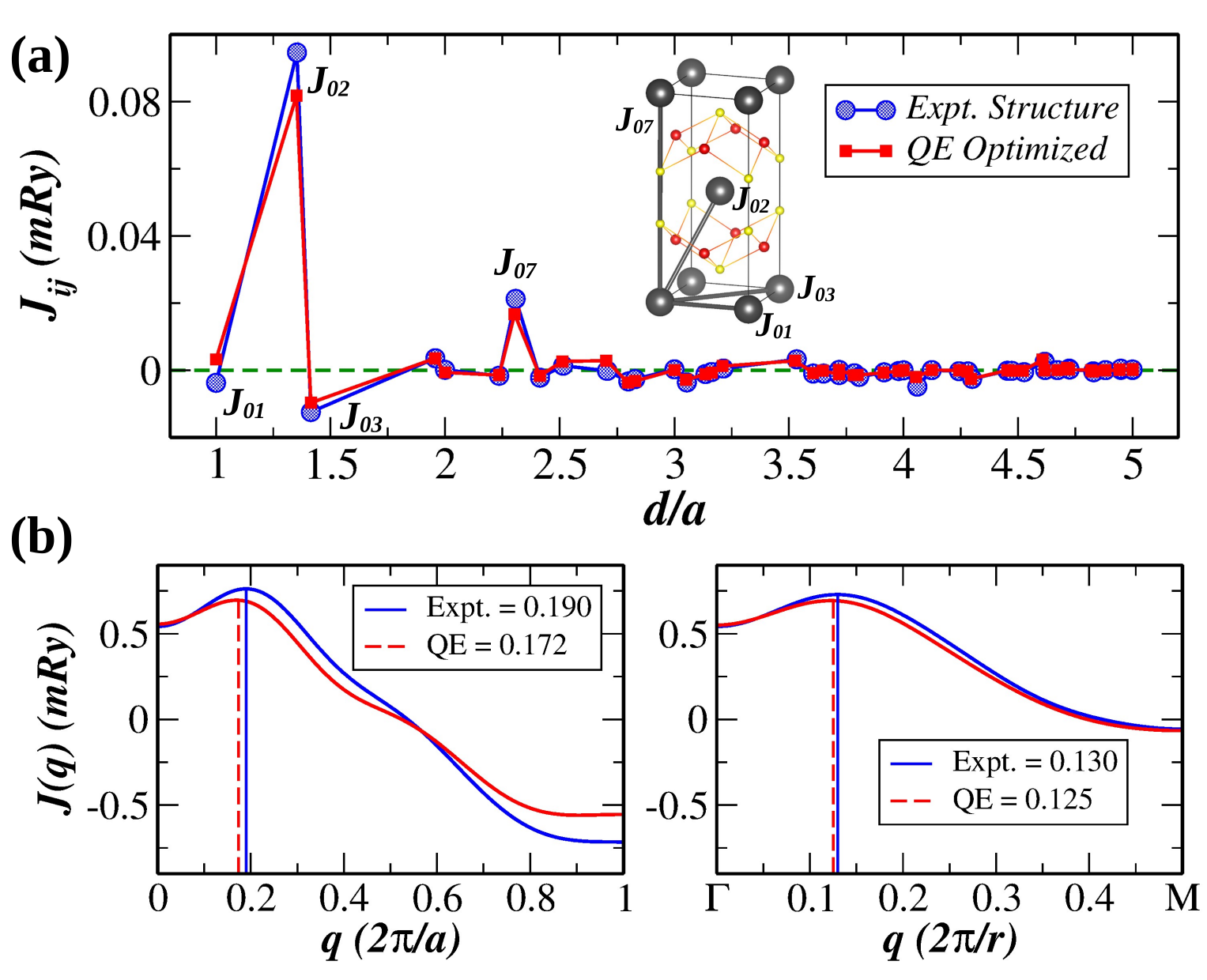}
\caption{A one-to-one comparison between the magnetic properties of the experimental structure and the QE optimized unstrained structure. (a) Calculated interatomic magnetic exchange interactions as a function of distance scaled by the lattice constant $a$. (b) Fourier transform of the same along $\Gamma - X - \Gamma$ (left) and $\Gamma - M$ (right) directions, respectively.}
\label{QE-magnetic}
\end{figure}
\FloatBarrier

\subsubsection{Magnetic Equivalence of Unstrained Structures}

We begin by comparing the magnetic properties of our computationally optimized unstrained structure with the experimental structure, as there are minor quantitative differences in their lattice parameters and bond lengths (Table~\ref{tab:str_table}). A one-to-one comparison of the calculated exchange interactions $J_{ij}$, using the same computational settings for both structures, and their Fourier transforms ($J(q)$) is shown in Fig.~\ref{QE-magnetic}. Only the first and second nearest-neighbor exchange parameters, $J_{01}$ and $J_{02}$, show some noticeable differences (Fig.~\ref{QE-magnetic} (a)). This is expected due to their stronger sensitivity to structural details, a point that we will explore when discussing the effects of strain (Fig.~\ref{exchange-change} (a)). However, these variations do not alter the qualitative features of $J(q)$ (Fig.~\ref{QE-magnetic} (b)) or the resulting spiral vectors, which determine the system's overall magnetic properties. For example, the calculated spiral vector $\vec Q_{100}$ along the $\Gamma - X$ direction is 0.172a* for the optimized structure, slightly less than the 0.190a* value from the experimental structure. This deviation only alters the periodicity of the spiral phases in the two-dimensional Gd layers, while their nature and ordering remain unchanged. Similar small changes are also observed for $\vec Q_{110}$ along $\Gamma - M$, which is a close energetic competitor to $\vec Q_{100}$ and becomes important only under strain.  Furthermore, the uniaxial anisotropy energies ($K_U$) for both structures are found to be very close ($~\sim$ 0.052 (0.057) meV / Gd for the experimental (optimized) structures). Given these results, we conclude that the optimized and experimental unstrained structures are magnetically equivalent. This important equivalence enables us to utilize our optimized structures to present data on the effect of strain, with confidence that the results accurately represent the real system.

\subsubsection{Effect of Uniaxial Strain on Magnetic Exchange}

\begin{figure}[ht]
\includegraphics[scale=0.5]{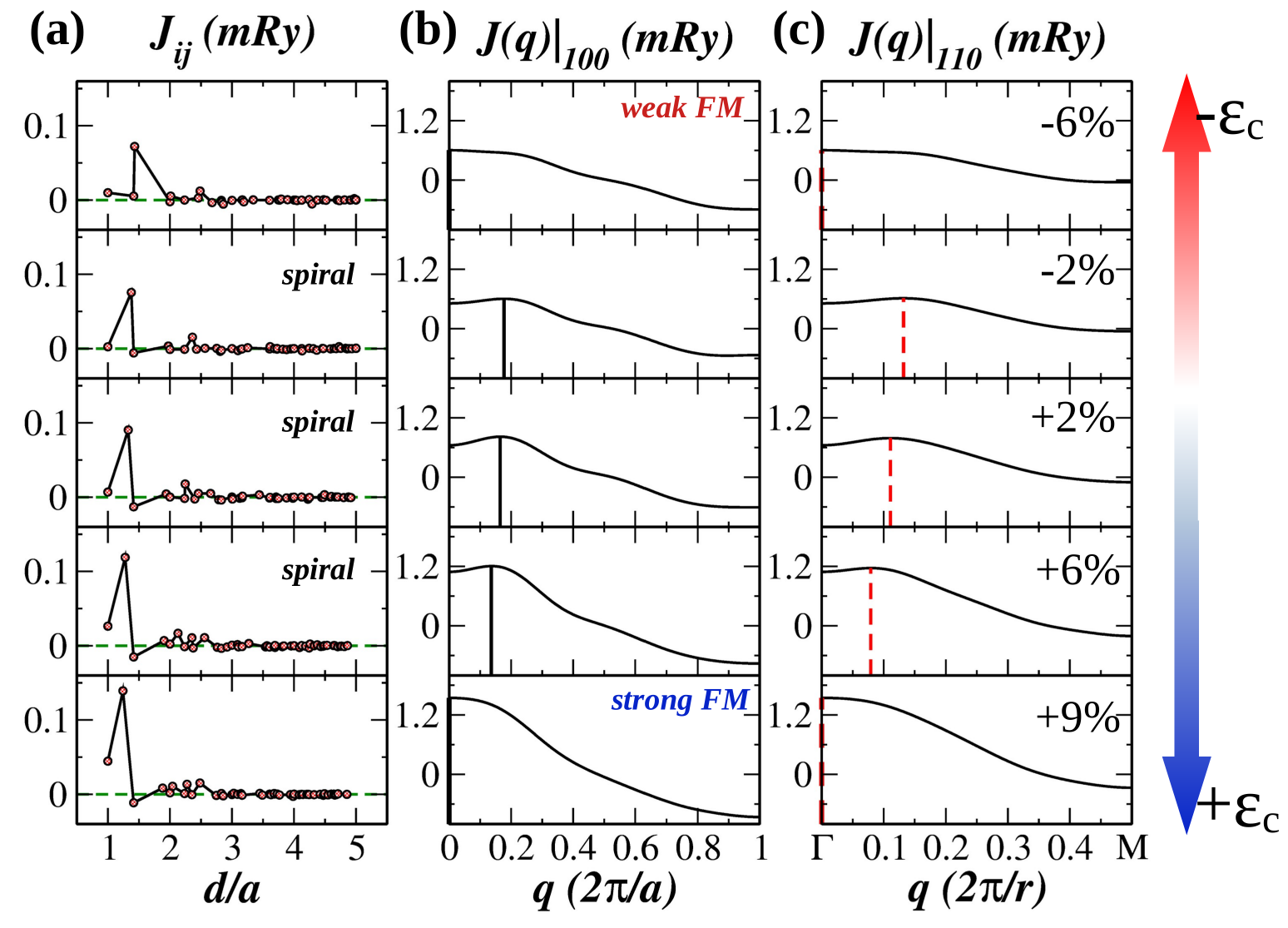}
\caption{(a) Variation in the calculated interatomic magnetic exchange interactions from tensile (-) to compressive values (+) of the uniaxial strain ($\epsilon_c$). The corresponding variations in the Fourier transform $J(q)$ are also shown along (b) $\Gamma - X - \Gamma$ and (c) $\Gamma - M$ directions, respectively. The peaks corresponding to the spiral modulation vectors $\vec Q_{100}$ and $\vec Q_{110}$ in (b) and (c) are depicted by solid/black and dashed/red vertical lines, respectively.} 
\label{exchange-change}
\end{figure}

The application of uniaxial strain ($\epsilon_c$) along the $c$ axis exerts a profound influence on the underlying magnetic exchange network, driving the system through several distinct magnetic regimes. Tracking and interpreting the strain-induced modulation of individual real-space exchange parameters, $J_{ij}$ (Fig.~\ref{exchange-change} (a), is inherently difficult due to the long-range oscillatory nature of the RKKY interaction. Therefore, we primarily rely on the Fourier transform $J(q)$ of $J_{ij}(R_{ij})$, which effectively encapsulates the effects of collective exchange in reciprocal space. Figs.~\ref{exchange-change} (b) and (c) present the calculated $J(q)$ along the high-symmetry $\Gamma-X$ and $\Gamma-M$ directions, respectively, for systems throughout the studied range of tensile and compressive strains. The positions of the peaks in $J(q)$, indicated by vertical lines, directly define the two most stable spiral modulation vectors, $\vec Q_{100}$ and $\vec Q_{110}$. An important observation is that extreme tensile ($\epsilon_c = -6\%$) and compressive ($\epsilon_c = +9\%$) strains completely suppress all exchange frustration, shifting the maximum of $J(q)$ to the $\Gamma$ point ($q=0$) and thus establishing a uniform FM order. However, these two FM states exhibit different stabilities. The compressive limit produces a strong FM (exchange energy of 1.55 mRy/Gd), a finding also qualitatively supported by the increasing magnitude of the strongest FM interactions, $J_{02}$ and $J_{01}$, under compression (see Fig.~\ref{exchange-change} (a)). In contrast, the tensile limit results in a weak FM, which has a much lower stabilization energy of only 0.61 mRy/Gd, which reflects an overall weakening of the exchange interactions. In all intermediate strain regimes, including the unstrained material, the finite values of $\vec Q_{100}$ and $\vec Q_{110}$ confirm the persistence of the exchange frustration and spiral magnetic phases. Hence, uniaxial strain serves as a crucial mechanism to modulate the observed magnetic phases.

\begin{figure}[ht]
\includegraphics[scale=0.5]{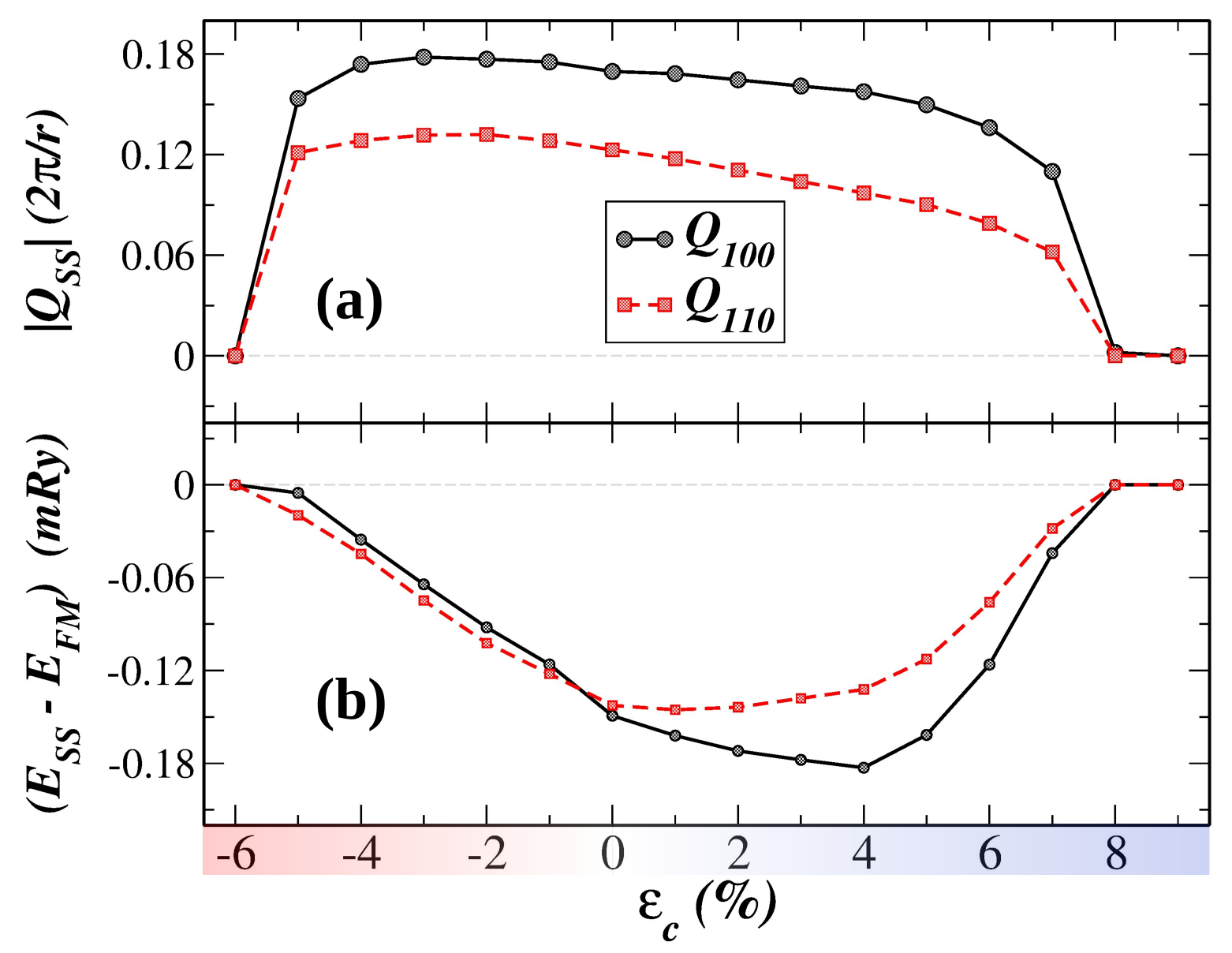}
\caption{(a) Variation in the magnitude of the spiral modulation vectors $\vec Q_{100}$ and $\vec Q_{110}$ from tensile (-) to compressive values (+) of the uniaxial strain ($\epsilon_c$). Values of $\vec Q_{100}$ and $\vec Q_{110}$ are expressed in the unit of the reciprocal lattice vectors $\vec G_{100} (2\pi/\vec a)$ and $\vec G_{110} (2\pi/\vec R_{110})$ respectively. (b) The corresponding variation in the stabilization energy (per Gd moment) of the spiral phases governed by $\vec Q_{100}$ and $\vec Q_{110}$ with respect to a uniform FM state.} 
\label{Jq-data}
\end{figure}
\FloatBarrier

To fully understand the evolution of the magnetic order, Fig.~\ref{Jq-data} illustrates the variation in the magnitude of the spiral vectors $\vec Q_{100}$ and $\vec Q_{110}$ alongside their energetic stability. The compressive strain causes the magnitude of both spiral $\vec{Q}$ vectors to decrease, suggesting the formation of longer-period spiral phases. This trend sharply terminates as the magnitude approaches zero when the system collapses into the strong FM state at 8\% compressive strain. Conversely, tensile strain leads to an increase in the magnitude of both $\vec{Q}$ vectors, implying a shorter-period spiral, before the sharp transition to the weak FM state at 6\% tensile strain. Further insight into the magnetic structure is revealed by the stabilization energies of the $\vec Q_{100}$ and $\vec Q_{110}$ spirals relative to the uniform FM state (Fig.~\ref{Jq-data} (b)). Under moderate compressive strain, the stability of the $\vec Q_{100}$-governed spiral initially increases with respect to both the FM state and the competitor $\vec Q_{110}$, reaching a maximum stabilization around $\epsilon_c = +4\%$. This enhanced stability, coupled with the shorter periodicity, suggests that moderate compression could be an effective way to stabilize the experimentally observed spiral phases with shorter periods. Thus, this could become a pathway toward stabilizing smaller-sized, more stable (against external magnetic fields) skyrmion structures. This prediction will be directly verified from the magnetic response of these particular systems under an external field and will be discussed later. Tensile strain presents a more complex landscape, inducing an energetic crossover that makes the $\vec Q_{110}$-governed spiral slightly more stable than the $\vec Q_{100}$ phase. Since the stabilization energies for $\vec Q_{100}$ and $\vec Q_{110}$ are very close in this tensile region, the possibility of a mixed magnetic order or a subtle transition to a $\vec Q_{110}$-driven ground state cannot be ignored. The shift toward $\vec Q_{110}$ is particularly interesting because it suggests the possibility of stabilizing a hexagonal skyrmion lattice, even though the magnetic Gd atoms are arranged on a fundamental square lattice. This complex behavior under tension is a clear manifestation of the significant weakening of the overall underlying exchange interactions and a drastic change in the underlying exchange frustration.

\subsubsection{Effect of Uniaxial Strain on Magnetic Anisotropy Energy}

\begin{figure}[ht]
    \centering
    \includegraphics[scale=0.50]{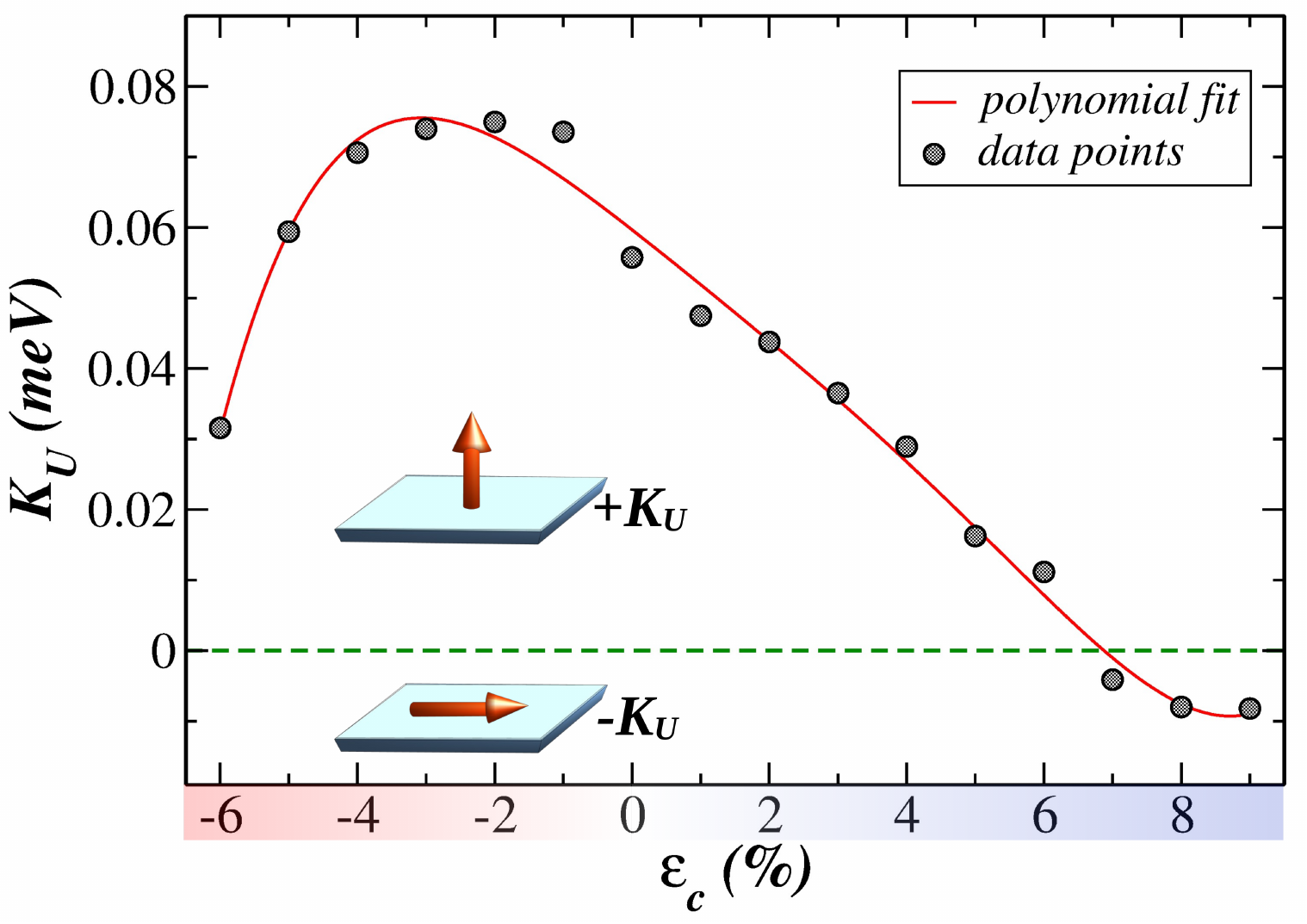}
    \caption{Variation of uniaxial anisotropy energy ($K_U$) as a function of the uniaxial strain ($\epsilon_c$). The solid line is a polynomial fit used here as a guide to the eye. The red arrows represent Gd moments that want to align perpendicular/parallel to the 2D layers for positive/negative values of $K_U$.}
    \label{ku_starin}
\end{figure}
\FloatBarrier

Beyond exchange interactions, the magnetic anisotropy energy (MAE) is a crucial factor in understanding the magnetic properties and the system's response to an external magnetic field~\cite{bouaziz2022fermi, TNomoto_JAP.133_2023, SSarkar_PRB.112_2025}. In most rare earth systems, MAE primarily comes from the strong spin-orbit coupling (SOC) of the highly anisotropic $4f$ charge cloud interacting with the crystalline electric field (CEF), which creates a large, on-site uniaxial anisotropy ($K_U$)~\cite{JJensen_OUP_1991}. However, Gd systems, which contain the $Gd^{3+}$ ion in the orbital angular momentum $L=0$ state, are a major exception because they do not have this intrinsic, single-ion anisotropy. Consequently, the MAE is much smaller, similar in size to $3d$ transition metal systems, and is instead mainly due to an indirect contribution from the conduction electrons~\cite{MCTosti_PRB.91_2003}. In this mechanism, the Gd $4f$ moment strongly polarizes the $5d$ conduction electrons through the $4f-5d$ intra-atomic Hund's exchange. Because of this, the MAE that originates from the SOC of the conduction electrons is effectively transferred back to the $4f$ spins~\cite{MCTosti_PRB.91_2003}. Our calculated $K_U$ value for GdRu$_2$Si$_2$ using the force theorem on the unstrained structure is about $+0.05$ meV/Gd atom, which is of the order of magnitude that we expect for Gd compounds. The positive sign here means an easy-axis anisotropy (out-of-plane $\vec \mu_{Gd}$, see Fig.~\ref{ku_starin} inset). This specific mechanism for MAE implies that uniaxial strain could modify the $K_U$ value by altering the electronic structure and properties of $5d$ conduction electrons. As we detail in the electronic properties section, our results show that the uniaxial strain significantly changes the electronic structure, causing the density of states at the Fermi energy to decrease or increase under compressive and tensile strain, respectively. As a result, changes in the band structure and Fermi surface, and hence in $K_U$, are expected. 

Fig.~\ref{ku_starin} illustrates the evolution of $K_U$ under strain, revealing an interesting correlation with the exchange frustration analysis discussed previously. With increasing compressive strain, $K_U$ gradually decreases in magnitude, similar to spiral vectors ($\vec Q$) observed under the same strain. Interestingly, at large compressive strains of $\epsilon_c \geq +8\%$, where the exchange analysis suggested a strong FM ground state (due to $\vec Q \rightarrow 0$), the anisotropy switches sign, becoming small and negative ($K_U \simeq -0.01$ meV/Gd). The negative sign implies an easy-plane (in-plane) orientation for the Gd moments (see Fig.~\ref{ku_starin} inset). However, the minute magnitude of $K_U$ suggests that this strong FM state is highly vulnerable to thermal fluctuations. Thus, this state is expected to behave nearly as a superparamagnetic state under an applied field, a finding confirmed by our ASD simulations detailed in the following section. In contrast, under tensile strain, the positive $K_U$ initially increases in magnitude up to $\epsilon_c \sim -2\%$, and then begins to decrease gradually. At the extreme tensile limit ($\epsilon_c = -6\%$), where the system transitions to a weak FM state, the anisotropy remains positive ($K_U \simeq +0.03$ meV/Gd). This positive value indicates that the moments in this weak FM ground state should retain their out-of-plane orientation, which we denote as $FM^Z$. The variation of $K_U$ with strain highlights its sensitivity to structural changes, underscoring its crucial role alongside exchange interactions in determining the stability of various phases and the magnetic response of the system. 

\subsubsection{Effect of Uniaxial Strain on Magnetic Response}

Now that we have the magnetic exchange ($J_{ij}$) and anisotropy ($K_U$) energies calculated for every strain value, we use $\text{ASD}$ simulations to model the system's magnetic response under an external field. This is a crucial final step to check how the magnetic phases are affected under strain.

\begin{figure}[ht]
    \centering
    \includegraphics[scale=0.50]{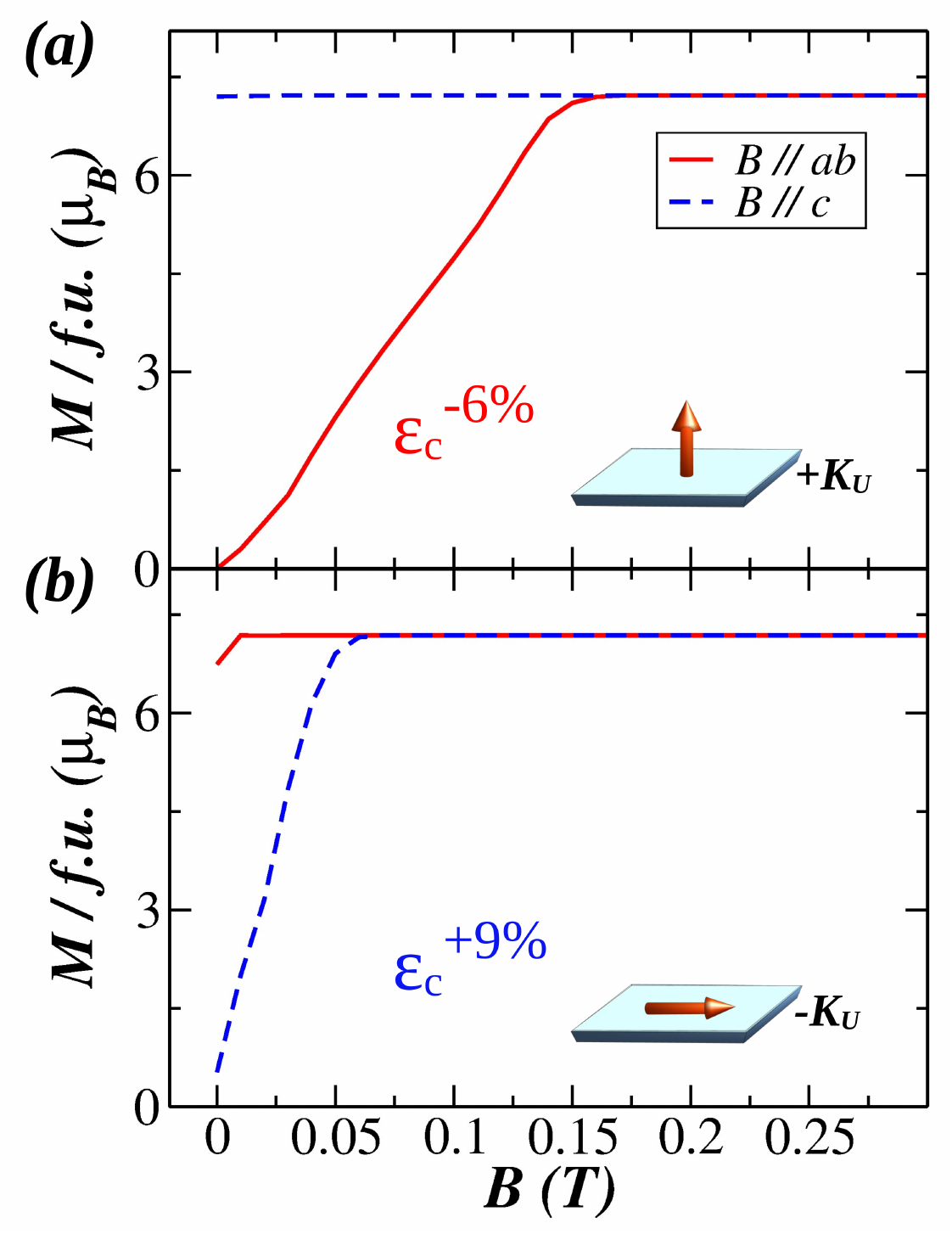}
    \caption{The ASD simulated $M(B)$ response with the calculated $J_{ij}$ and $K_U$ for strained systems with uniaxial strain ($\epsilon_c$) of (a) -6 \% \textbf{(tensile strain)} and (b) +9 \% \textbf{(compressive strain)}, where $J_{ij}$ suggests a FM state, but are with different types of $K_U$ as indicated in the insets. See text for more details.}
    \label{MB-FM_response}
\end{figure}

We first looked at the extreme limits of strain, where our earlier analysis suggested a strong $\text{FM}$ state under compressive strain ($\epsilon_c = +9 \%$) and a weak $\text{FM}$ state under tensile strain ($\epsilon_c = -6 \%$). The simulated magnetization curves, $M(B)$, for these two extreme cases (Fig.~\ref{MB-FM_response}) show exactly what we expect: an in-plane $\text{FM}$ response for the compressive limit (due to easy-plane anisotropy) and an out-of-plane $\text{FM}$ response for the tensile limit (due to easy-axis anisotropy). Interestingly, despite the stronger exchange interactions in the extreme compressive case, due to small easy-plane anisotropy (see Fig.~\ref{ku_starin}), the system quickly field polarizes along the $c$-axis at a small field of just $0.07~ \text{T}$, almost behaving like a superparamagnet. In contrast, the weak $\text{FM}$ state at the extreme tensile limit, stabilized by easy-axis anisotropy, requires a much larger in-plane critical field of $\sim 0.20~\text{T}$ to fully field-polarize the moments.

\begin{figure}[ht]
    \centering
    \includegraphics[scale=0.60]{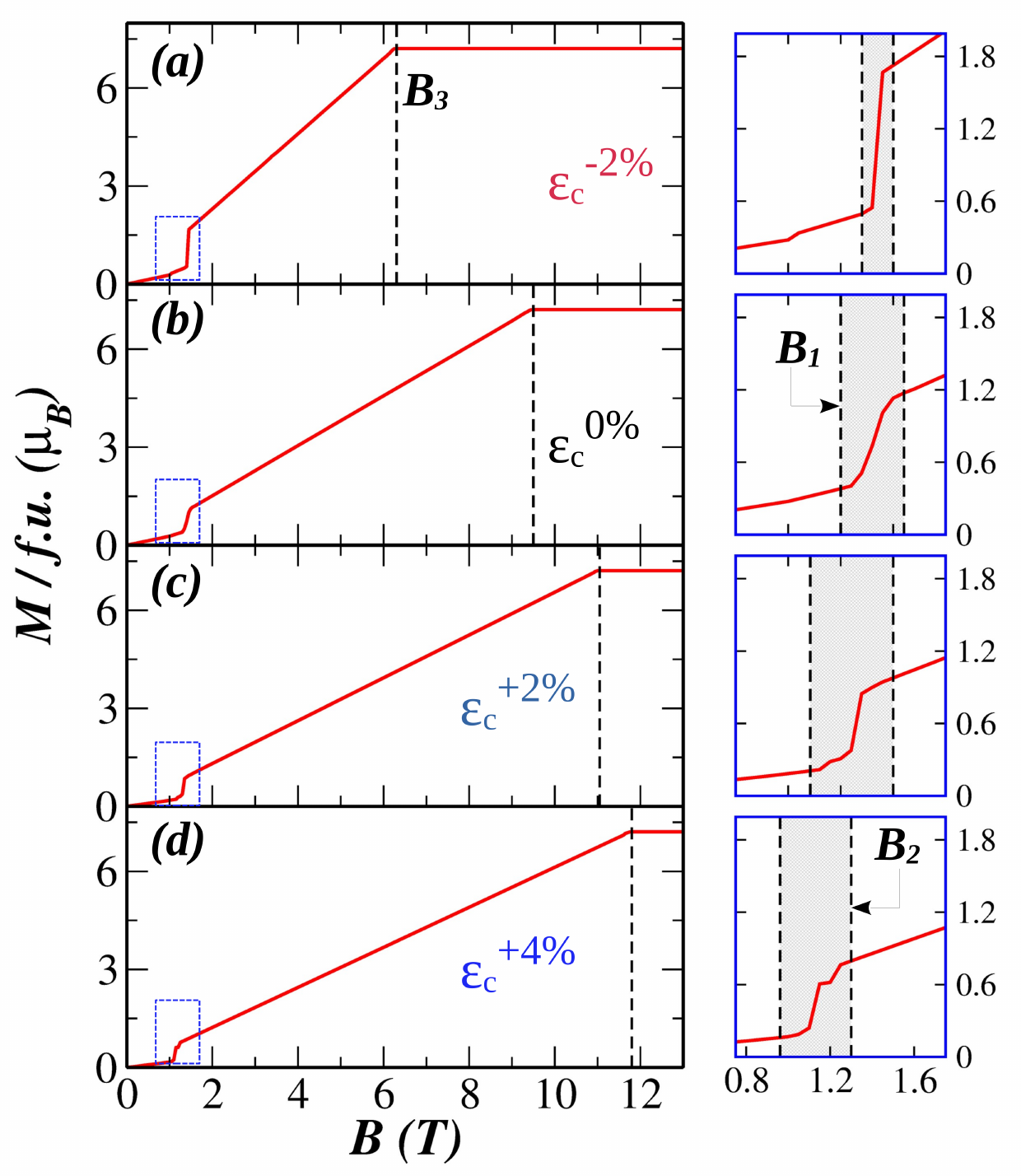}
    \caption{The ASD simulated $M(B)$ response with the calculated $J_{ij}$ and $K_U$ for some strained systems with uniaxial strain ($\epsilon_c$) of (a) -2 \%, (b) 0 \%, (c) +2 \%, and (d) +4 \%. On the right side boxes in each subfigure, a magnified view of the enclosed area inside the box with dotted lines is shown. The vertical dashed lines represent the three critical fields where the system undergoes a magnetic phase transition. See text for more details.}
    \label{MB_response}
\end{figure}
\FloatBarrier

For the intermediate strain values, where we see nonzero spiral vectors $\vec Q_{100}$ and $\vec Q_{110}$, the system undergoes a series of transitions under an applied field $B||c$ before reaching the final field-polarized ($\text{FP}$) FM state denoted as $\text{FM}^Z$. The simulated $M(B)$ response for some of the strained systems, including the unstrained system, is shown in Fig.~\ref{MB_response}. Here, three critical fields ($B_1, B_2,$ and $B_3$) as indicated with dotted lines mark the transitions from Phase I to II, II to III, and III to $\text{FM}^Z$, respectively~\cite{khanh2020nanometric}. These critical field values were determined from the peak positions and widths in the $dM/dB$ vs $B$ data, with further details provided in Section~II of the SM~\cite{sm_sarkar2025}.  The shaded region between $B_1$ and $B_2$, as shown on the right side of Fig.~\ref{MB_response}, corresponds to Phase II, which is experimentally known to host the Skyrmion Lattice ($\text{SkL}$) state~\cite{khanh2020nanometric}. What we found here is threefold: 1) $B_3$ gradually increases with compressive strain, while the tensile strain lowers it. 2) Phase II starts to appear in slightly lower fields with compressive strain, but in higher fields with tensile strain. 3) The width of the shaded region (persistence of Phase II) initially increases slightly under compression before starting to destabilize again. This enhancement of the $\text{SkL}$ phase under moderate uniaxial compression is consistent with recent experimental work by L. Gries \textit{et al.}~\cite{LGries_PRB.111_2025}, where increased stability of the SkL phase was achieved with uniaxial pressure.

\begin{figure}[ht]
\includegraphics[scale=0.50]{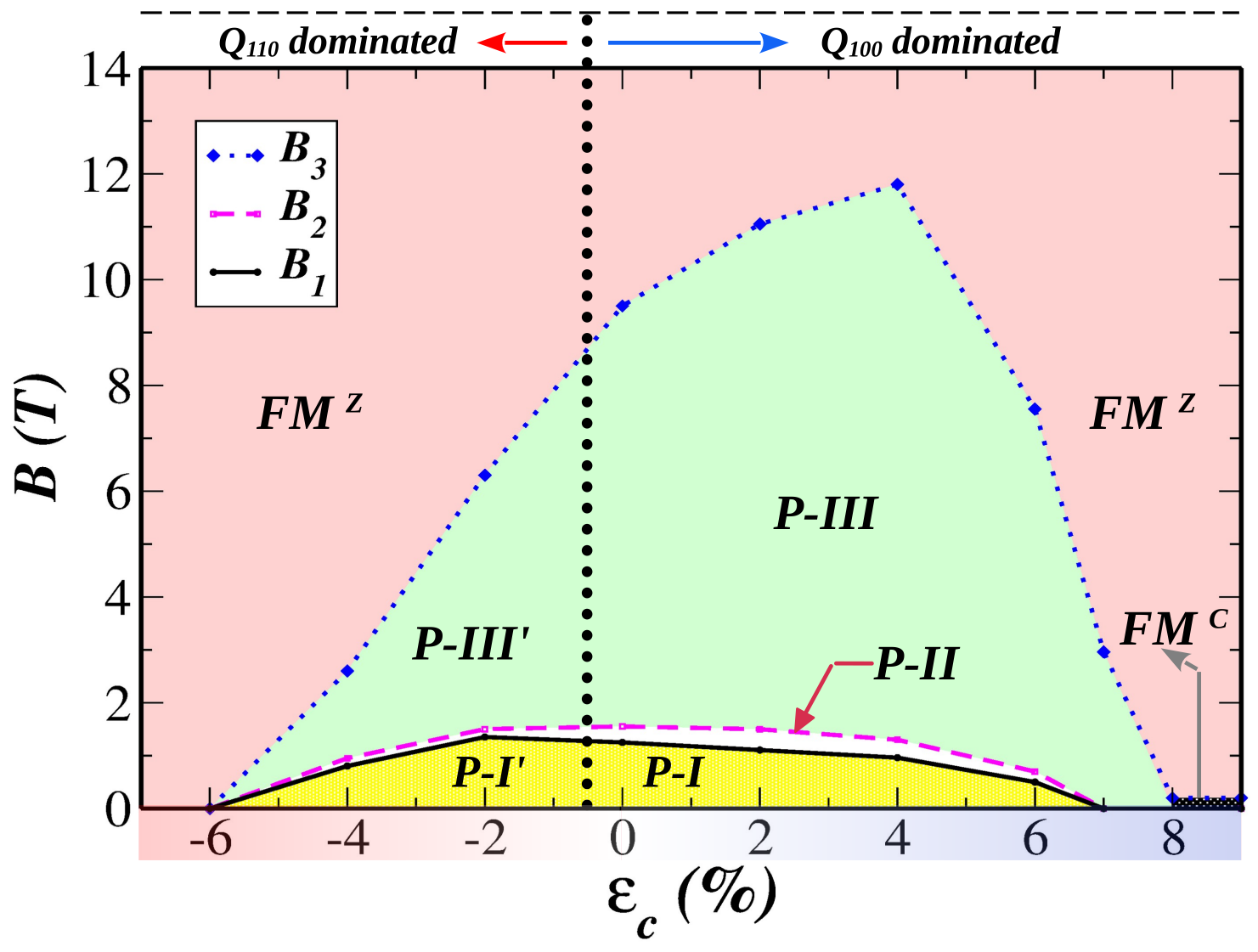}
\caption{ $B$ vs $\epsilon_c$ phase diagram, constructed from the variation of the critical fields $B_1$, $B_2$, and $B_3$. This shows different magnetic phases and their transformation from one to another as a function of $B$ and $\epsilon_c$. The white narrow region between Phase I ($\text{I}'$) and III ($\text{III}'$) is Phase II ($\text{II}'$). $\text{FM}^Z$ means a field-polarized FM state along the $c$ axis. $\text{FM}^C$ means a canted FM state. See text for more details.}
\label{phase-diagram}
\end{figure}
\FloatBarrier

To better understand the connection between our microscopic parameters ($J_{ij}$, $K_U$) and the bulk response, we plotted the three critical fields ($B_1, B_2, B_3$) as a function of strain to construct a full magnetic phase diagram (Fig.~\ref{phase-diagram}). This phase diagram makes important revelations: The stability of Phase I is directly related to the uniaxial anisotropy energy $K_U$. The critical field $B_1$, where Phase I disappears, increases or decreases following the exact trend of $K_U$ (Fig.~\ref{ku_starin}). Similarly, the stability of Phase III, measured by $B_3$, is directly controlled by the energetic stability of the spiral vectors ($Q_{100}/Q_{110}$), as shown in Fig.~\ref{Jq-data}. For example, the energetic stability of $\vec Q_{100}$ with respect to a $\text{FM}$ state increases until $+4\%$ compressive strain, and Phase III shows an identical trend, with $B_3$ rising from $9.50~\text{T}$ to $11.80~\text{T}$ before gradually falling. Notably, at $7\%$ compression, $K_U$ is nearly zero, and phases I and II do not occur; Phase III becomes the zero-field ground state due to the weak exchange frustration that remains. Further compression into the $8-9\%$ range results in an $\text{FM}$ ground state with in-plane spins (easy-plane $K_U$). Under an applied field $B||c$, this transforms into a canted $\text{FM}$ state ($\text{FM}^C$) for a very narrow field range (indicated by a small black-shaded region in Fig.~\ref{phase-diagram}), before fully polarizing into the $\text{FM}^z$ state.

The overall phase diagram (Fig.~\ref{phase-diagram}) demonstrates the robust tunability of the magnetic response in GdRu$_2$Si$_2$ through mechanical strain. Specifically, under tensile strain, the relative energetic stability of the $\vec Q_{110}$ wave vector is enhanced over $\vec Q_{100}$ (see Fig.~\ref{Jq-data}), leading to the emergence of the distinct phases $\text{I}'$, $\text{II}'$, and $\text{III}'$ compared to the known phases $\text{I}$, $\text{II}$, and $\text{III}$. The spin structure factor and magnetic ordering for all phases appearing in the phase diagram (Fig.~\ref{phase-diagram}) can be found in Section~IV of the SM~\cite{sm_sarkar2025}. It is important to briefly clarify the role of additional interactions. Although the fundamental magnetic response, phase transitions, and the very existence of these phases are primarily governed by the intricate interplay between exchange coupling, uniaxial anisotropy, and the applied magnetic field~\cite{nomoto2020formation, TNomoto_JAP.133_2023, bouaziz2022fermi, SSarkar_PRB.112_2025}, fine-tuning of the final magnetic ordering may require the inclusion of weaker, sub-dominant terms. Specifically, the exact magnetic ordering in Phases I, II, and III found in recent experiments~\cite{NDKhanh_AdvSci.9_2022, GWood_PRB.107_2023, JSpethmann_PRM.8_2024} often requires the inclusion of weaker interactions, such as dipolar terms for Phases I and II (see our work~\cite{SSarkar_PRB.112_2025}), where in this study as well we see that dipolar interactions can assist the skyrmion stability and modify the other phases (the effect of dipolar interactions on spin configuration is shown in the supplementary information), but this is a topic for a more detailed analysis in a future work, and biquadratic exchange for field-induced $\text{SkL}$ (Phase II) and Meron-like phase (Phase III)~\cite{hayami2021square}. Our focus here is on the dominant exchange-anisotropy-field competition, which is strongly affected by strain, thereby providing the foundation for the observed phase diagram.

\section{Conclusions}

In summary, our work offers a deeper understanding of the correlation between structural degrees of freedom and magnetic phase stability in GdRu$_2$Si$_2$, showing an energy-efficient route for manipulating its exotic magnetic textures. By combining first-principles density functional theory calculations with classical spin model simulations, we meticulously demonstrated that the exchange, anisotropy, and magnetic phases are exquisitely sensitive to strain. Our analysis reveals that compressive strain up to $\sim 2.0\%$ alters the magnetic landscape, significantly favoring the $\vec Q_{100}$ magnetic ordering vector and, consequently, expanding the stability window of the topologically non-trivial field-induced phases in the magnetic phase diagram. In contrast, tensile strain induces a crossover to a $\vec Q_{110}$ ordering, signifying a completely different magnetic ground state and a distinct set of emergent field-induced phases. These findings not only deepen the theoretical understanding of GdRu$_2$Si$_2$'s magnetism but also provide direct and quantitative guidance for experimental strain engineering.

\section*{Acknowledgements}

This work was financially supported by the Knut and Alice Wallenberg Foundation through grant numbers 2018.0060, 2021.0246, and 2022.0108 (PI's: O.E. and A.D.).  R.P. and V.B. acknowledge further support from the G\"oran Gustafsson Foundation (recipient of the ``small prize'': V.B.). S.S. acknowledges funding from the Carl Tryggers Foundation (grant number CTS 22:2013, PI: V.B.). V.B. also acknowledges support from the Swedish Research Council through Grant No. 2024-05206 and the Ministry of Education, Youth and Sports of the Czech Republic through the e-INFRA CZ (ID:90254). O.E. and A.D. acknowledge support from the Wallenberg Initiative Materials Science for Sustainability (WISE) funded by the Knut and Alice Wallenberg Foundation (KAW). A.D. also acknowledges financial support from the Swedish Research Council (Vetenskapsrådet, VR), Grant No. 2016-05980, Grant No. 2019-05304, and Grant No. 2024-04986. O.E. also acknowledges support by the Swedish Research Council (VR), the Foundation for Strategic Research (SSF), the Swedish Energy Agency (Energimyndigheten), the European Research Council (854843-FASTCORR), eSSENCE and STandUP.

The computations/data handling were enabled by resources provided by the Swedish National Infrastructure for Computing (SNIC) at the National Supercomputing Centre (NSC, Tetralith cluster) partially funded by the Swedish Research Council through grant agreement no.\,2018-05973 and by the National Academic Infrastructure for Supercomputing in Sweden (NAISS) at the National Supercomputing Centre (NSC, Tetralith cluster) partially funded by the Swedish Research Council through grant agreement no.\,2022-06725. We acknowledge VSB – Technical University of Ostrava, IT4Innovations National Supercomputing Center, Czech Republic, for awarding this project access to the LUMI supercomputer, owned by the EuroHPC Joint Undertaking, hosted by CSC (Finland) and the LUMI consortium through the Ministry of Education, Youth and Sports of the Czech Republic through the e-INFRA CZ (grant ID: 90254). We also acknowledge the EuroHPC Joint Undertaking for awarding us access to Karolina supercomputer at IT4Innovations, Czech Republic. The structural sketches in Figs.~\ref{lattice-geo},~\ref{elf-fig}, and \ref{QE-magnetic} were produced using \textsc{VESTA3} software \cite{vesta}.

\bibliography{main}

\end{document}


\title{Supplementary material for Mechanical control of magnetic exchange and response in GdRu$_2$Si$_2$: A computational study}

\author{Sagar Sarkar}\thanks{These authors contributed equally to this work}
\affiliation{Department of Physics and Astronomy, Uppsala University, Uppsala, 751 20, Sweden.}

\author{Rohit Pathak}\thanks{These authors contributed equally to this work}
\affiliation{Department of Physics and Astronomy, Uppsala University, Uppsala, 751 20, Sweden.}

\author{Arnob Mukherjee}
\affiliation{Department of Physics and Astronomy, Uppsala University, Uppsala, 751 20, Sweden.}

\author{Anna Delin}
    \affiliation{\KTH}
    \affiliation{\WISEKTH}
    \affiliation{\SeRC}

\author{Olle Eriksson}
\affiliation{Department of Physics and Astronomy, Uppsala University, Uppsala, 751 20, Sweden.}
\affiliation{Wallenberg Initiative Materials Science for Sustainability, Uppsala University, 75121 Uppsala, Sweden.}

\author{Vladislav Borisov}
\affiliation{Department of Physics and Astronomy, Uppsala University, Uppsala, 751 20, Sweden.}

\date{\today}

\maketitle

\section{ASD simulation cell dimension}

In our recent work on GdRu$_2$Si$_2$~\cite{SSarkar_PRB.112_2025}, we established that it is essentially a 3D magnet and an ASD simulation cell dimension $N\times N \times N$ is required, where $N$ is some integer.  The value of $N$ should be chosen in such a way that the simulation box can accommodate an integer number of wavelengths of spiral states. This is important for the best outcome from the ASD simulation. We have performed ASD simulations to obtain the $M~ \text{vs}~B $ response for the strained systems with $\epsilon_c$ = $-6\%$, $-4\%$, $-2\%$, $0\%$, $+2\%$, $+4\%$, $+6\%$, $+7\%$, and $+9\%$ (see the main text for more details). Table~\ref{dimension_table} provides detailed information on the spiral wave vectors, wavelengths, and the determined value of $N$ for these simulations. 

\begin{table}[h!]
\caption{Theoretically obtained in-plane spiral wavelengths $\lambda_{100}$ and $\lambda_{110}$ for different strained systems considered for ASD simulations. The value of $N$ in the simulation cell dimensions ($N \times N \times N$) is given that accommodates $m$ number of spin spiral wave lengths $\lambda_{100}$ or $\lambda_{110}$.}
\label{dimension_table}
\begin{tabular}{|c|c|c|c|c|c|c|c|}
\hline
\begin{tabular}[c]{@{}c@{}}Strain \\ $\epsilon_c~(\%)$\end{tabular} &
  $|\vec Q_{100}|$ &
  $\lambda_{100}$ &
  $|\vec Q_{110}|$ &
  $\lambda_{110}$ &
  \begin{tabular}[c]{@{}c@{}}Dominating\\ $\vec Q$\end{tabular} &
  \begin{tabular}[c]{@{}c@{}}$m$\\ in\\ $(m \times \lambda)$\end{tabular} &
  \begin{tabular}[c]{@{}c@{}}Used\\ $N$\end{tabular} \\ \hline
-6 & 0      & NA     & 0      & NA     & NA    & NA & 35 \\ \hline
-4 & 0.1739 & 5.7500 & 0.2570 & 3.8913 & $\vec Q_{110}$ & 9  & 35 \\ \hline
-2 & 0.1769 & 5.6517 & 0.2640 & 3.7872 & $\vec Q_{110}$ & 9  & 34 \\ \hline
0  & 0.1697 & 5.8941 & 0.2458 & 4.0683 & $\vec Q_{100}$ & 6  & 35 \\ \hline
+2 & 0.1647 & 6.0732 & 0.2216 & 4.5126 & $\vec Q_{100}$ & 6  & 36 \\ \hline
+4 & 0.1576 & 6.3461 & 0.1943 & 5.1467 & $\vec Q_{100}$ & 6  & 38 \\ \hline
+6 & 0.1362 & 7.3433 & 0.1580 & 6.3291 & $\vec Q_{100}$ & 5  & 37 \\ \hline
+7 & 0.1100 & 9.0926 & 0.1239 & 8.0710 & $\vec Q_{100}$ & 4  & 36 \\ \hline
+9 & 0      & NA     & 0      & NA     & NA    & NA & 35 \\ \hline
\end{tabular}
\end{table}
\FloatBarrier

\section{Determination of critical fields $B_1$, $B_2$, and $B_3$}

\begin{figure}[ht]
\includegraphics[scale=0.55]{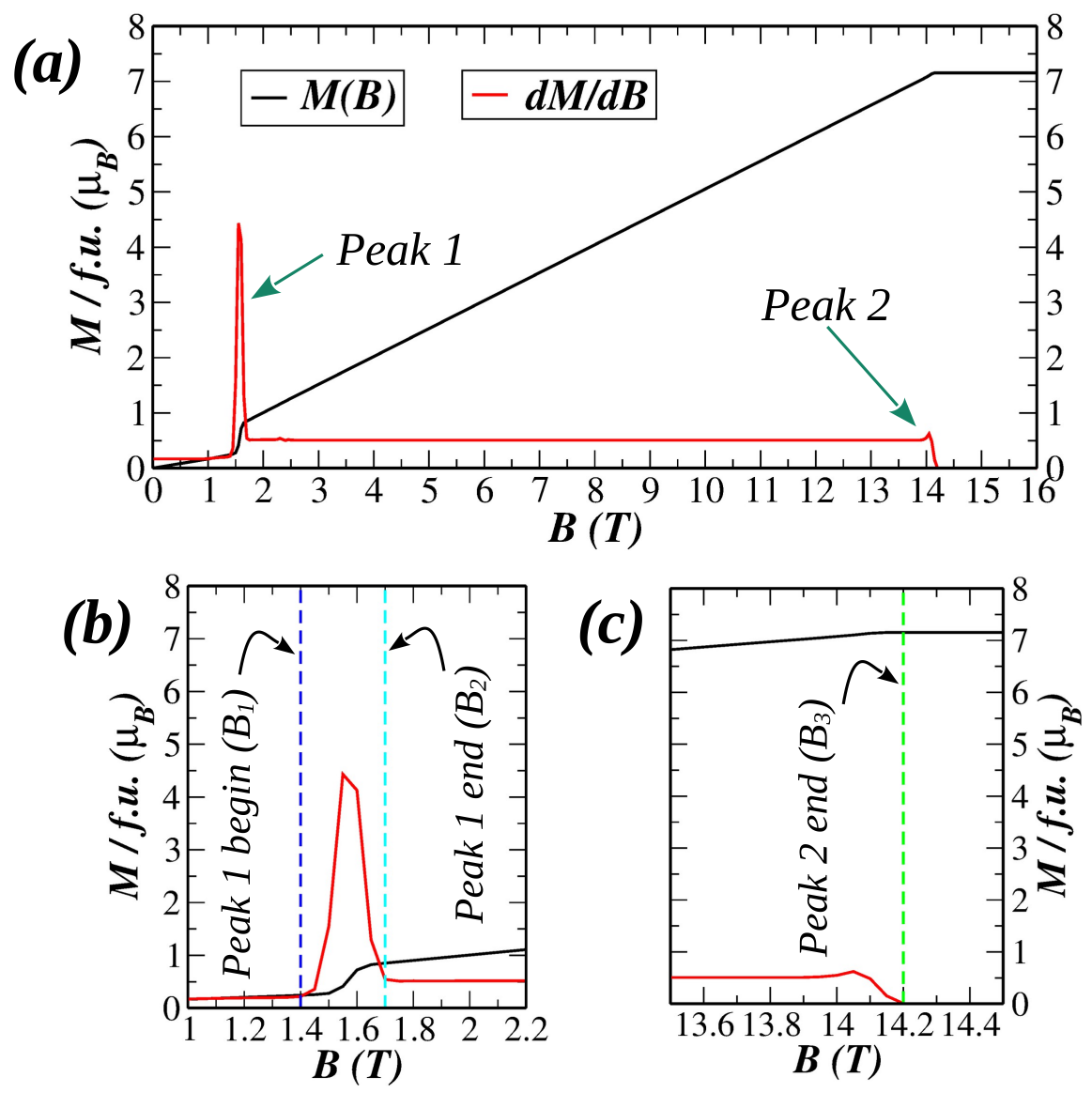}
\caption{(a) The simulated $M(B)$ response of GdRu$_2$Si$_2$ with experimental structure, and the corresponding $dM/dB$ data showing the position of two peaks. Simulation done with exchange ($J_{ij}$) and uniaxial anisotropy ($K_U$) in the spin Hamiltonian. (b) A magnified view of peak 1. The two dashed lines indicate the starting and ending of peak 1, and the corresponding magnetic fields are considered as the critical fields $B_1$ and $B_2$, respectively. (c) A magnified view of peak 1. The dashed line indicates the end of peak 2 and the critical field $B_3$.}
\label{derivative_expt}
\end{figure}

GdRu$_2$Si$_2$ undergoes interesting magnetic field-driven phase transitions consisting of different spiral phases~\cite{khanh2020nanometric}. In the $M(B)$ response, there are three critical fields $B_1$, $B_2$, and $B_3$, which mark the onset of phase transitions. At $B_1$, $B_2$, and $B_3$, a phase transition occurs from Phase (I to II), (II to III), and (III to FP-FM), respectively. To extract $B_1$, $B_2$, and $B_3$ from the $M(B)$ data, we calculate $dM/dB$ as a function of $B$, showing susceptibility peaks. As shown schematically in Fig.~\ref{derivative_expt}, we consider $B_1$ as the point where the first peak starts and $B_2$ as the point where it ends. $B_3$ is identified as the point where the second peak ends. 

\FloatBarrier

In our study, we performed $M(B)$ calculations for different strained systems with $\epsilon_c$ = $-4~\%$, $-2~\%$, $0~\%$, $+2~\%$, $+4~\%$, $+6~\%$, and $+7~\%$, showing spiral phases. The corresponding simulated $M(B)$ responses and the determined critical field values, as described above, are shown in the following figures.

\begin{figure}[ht]
\includegraphics[scale=0.5]{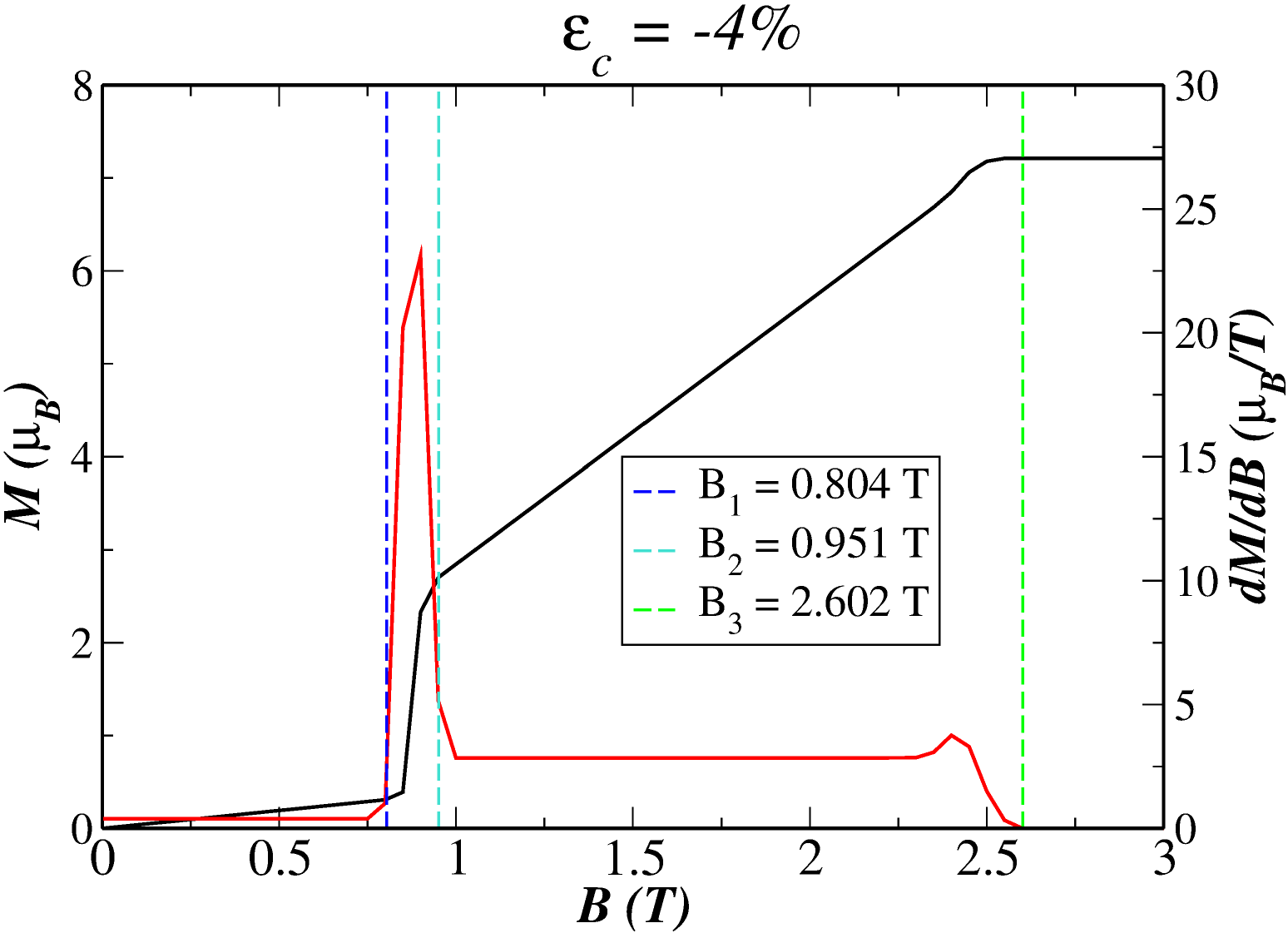}
\caption{System $\epsilon_c$ = $-4~\%$: The simulated $M(B)$ response and the corresponding $dM/dB$ data showing the position of peaks and determined critical fields $B_1$, $B_2$, and $B_3$.}
\label{derivative_pm4}
\end{figure}
\FloatBarrier

\begin{figure}[ht]
\includegraphics[scale=0.5]{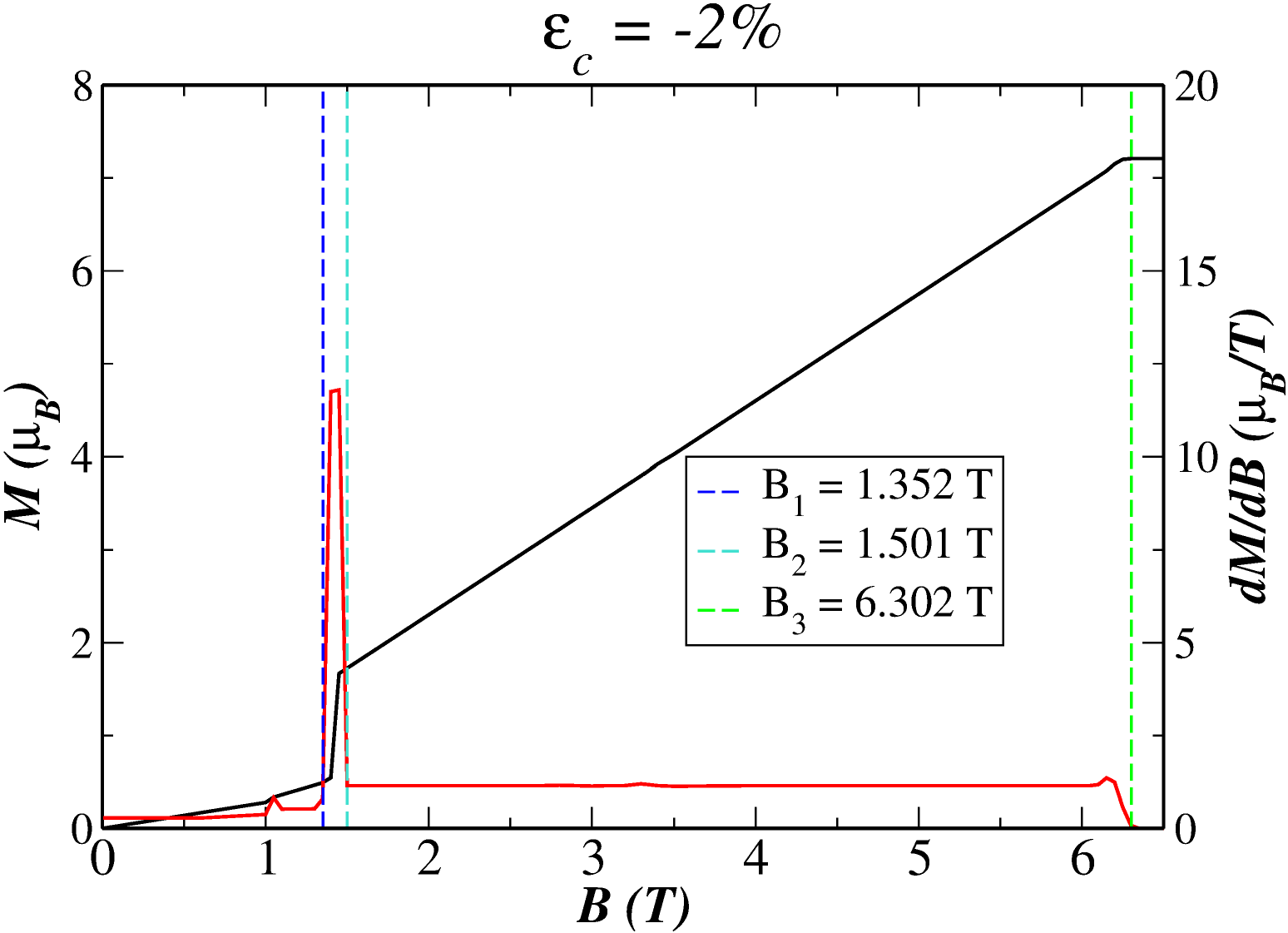}
\caption{System $\epsilon_c$ = $-2~\%$: The simulated $M(B)$ response and the corresponding $dM/dB$ data showing the position of peaks and determined critical fields $B_1$, $B_2$, and $B_3$.}
\label{derivative_pm2}
\end{figure}
\FloatBarrier

\begin{figure}[ht]
\includegraphics[scale=0.5]{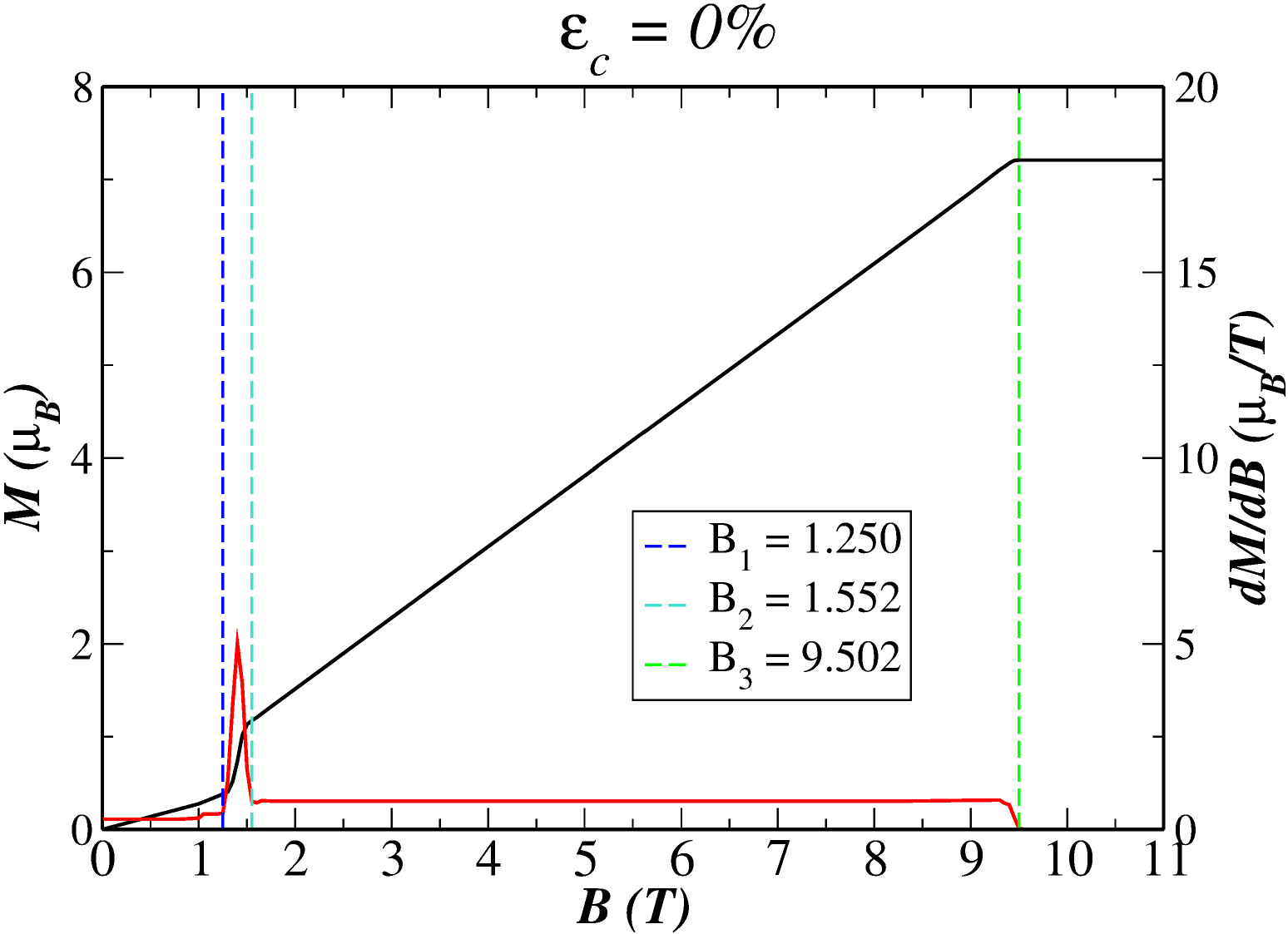}
\caption{System $\epsilon_c$ = $0~\%$: The simulated $M(B)$ response and the corresponding $dM/dB$ data showing the position of peaks and determined critical fields $B_1$, $B_2$, and $B_3$.}
\label{derivative_p0}
\end{figure}
\FloatBarrier

\begin{figure}[ht]
\includegraphics[scale=0.5]{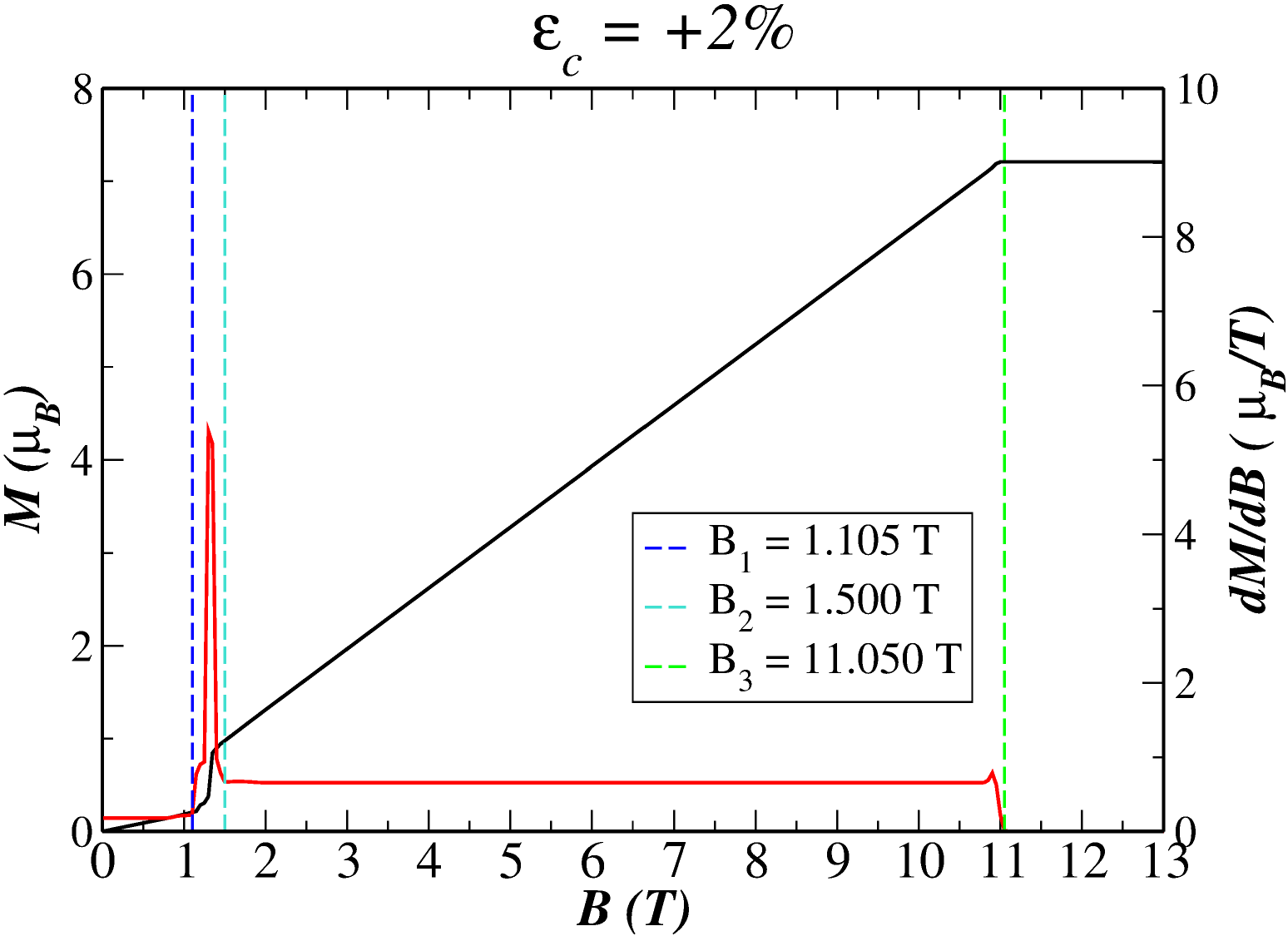}
\caption{System $\epsilon_c$ = $+2~\%$: The simulated $M(B)$ response and the corresponding $dM/dB$ data showing the position of peaks and determined critical fields $B_1$, $B_2$, and $B_3$.}
\label{derivative_p2}
\end{figure}
\FloatBarrier

\begin{figure}[ht]
\includegraphics[scale=0.5]{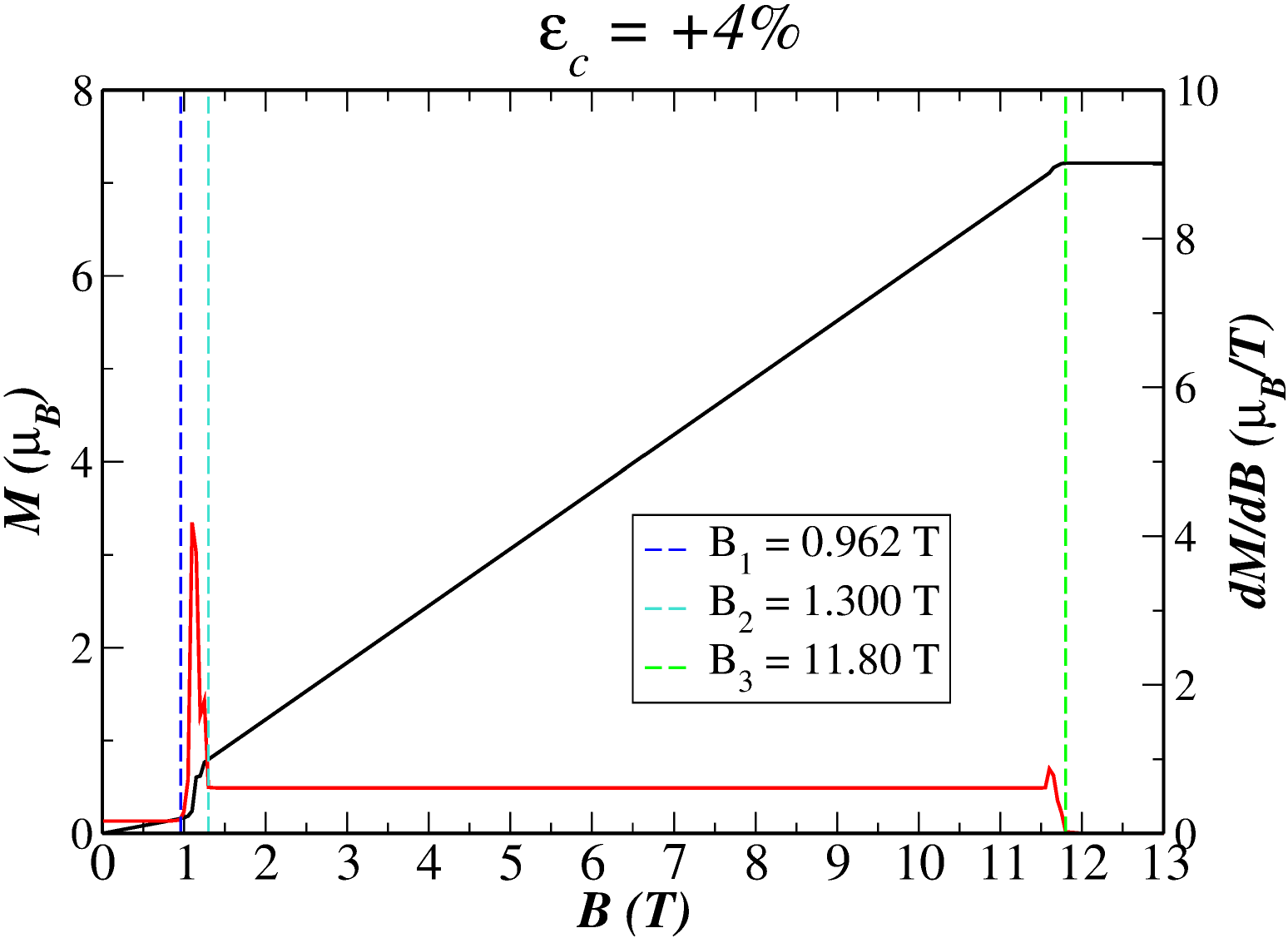}
\caption{System $\epsilon_c$ = $+4~\%$: The simulated $M(B)$ response and the corresponding $dM/dB$ data showing the position of peaks and determined critical fields $B_1$, $B_2$, and $B_3$.}
\label{derivative_p4}
\end{figure}
\FloatBarrier

\begin{figure}[ht]
\includegraphics[scale=0.5]{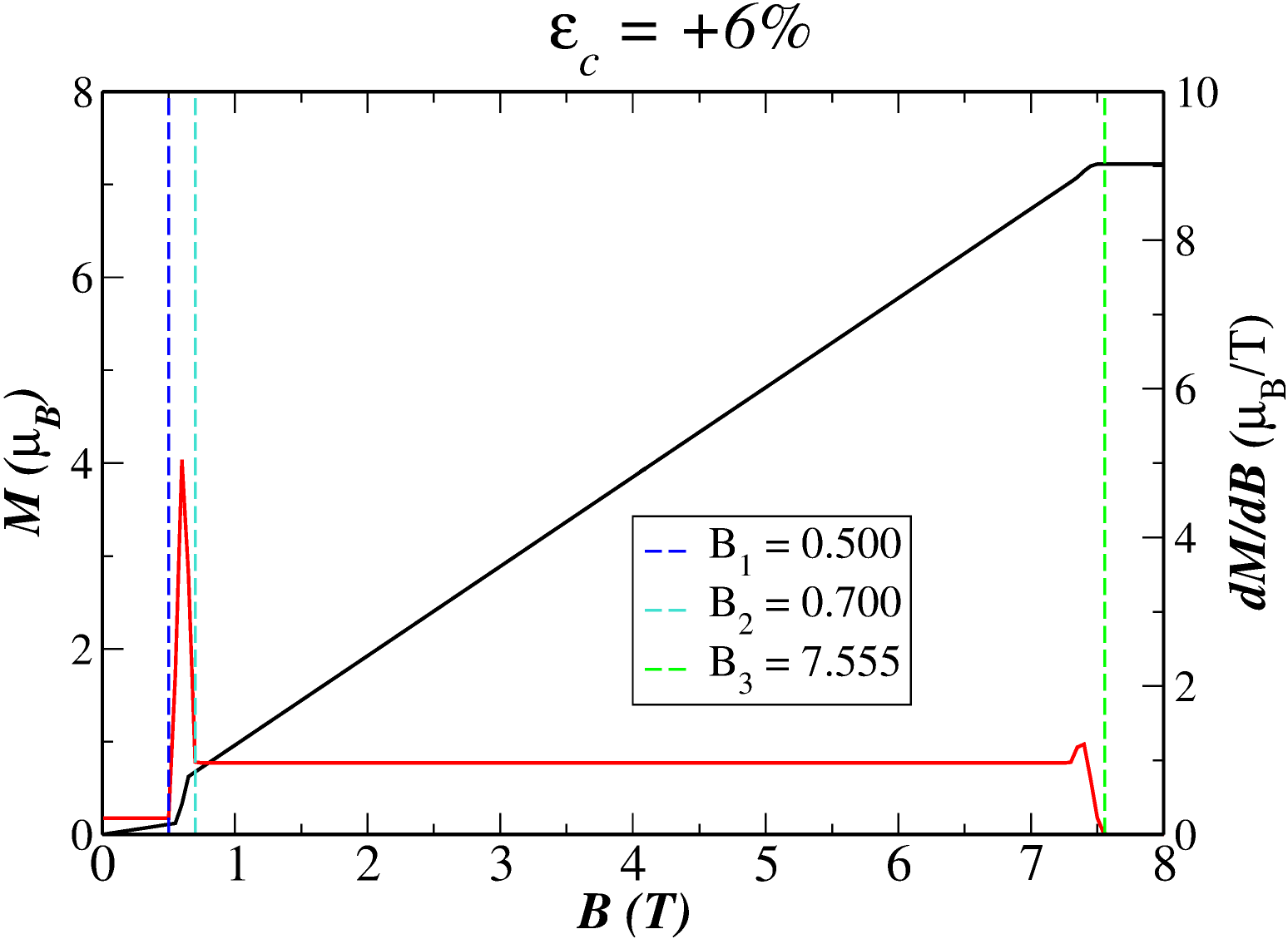}
\caption{System $\epsilon_c$ = $+6~\%$: The simulated $M(B)$ response and the corresponding $dM/dB$ data showing the position of peaks and determined critical fields $B_1$, $B_2$, and $B_3$.}
\label{derivative_p6}
\end{figure}
\FloatBarrier

\begin{figure}[ht]
\includegraphics[scale=0.5]{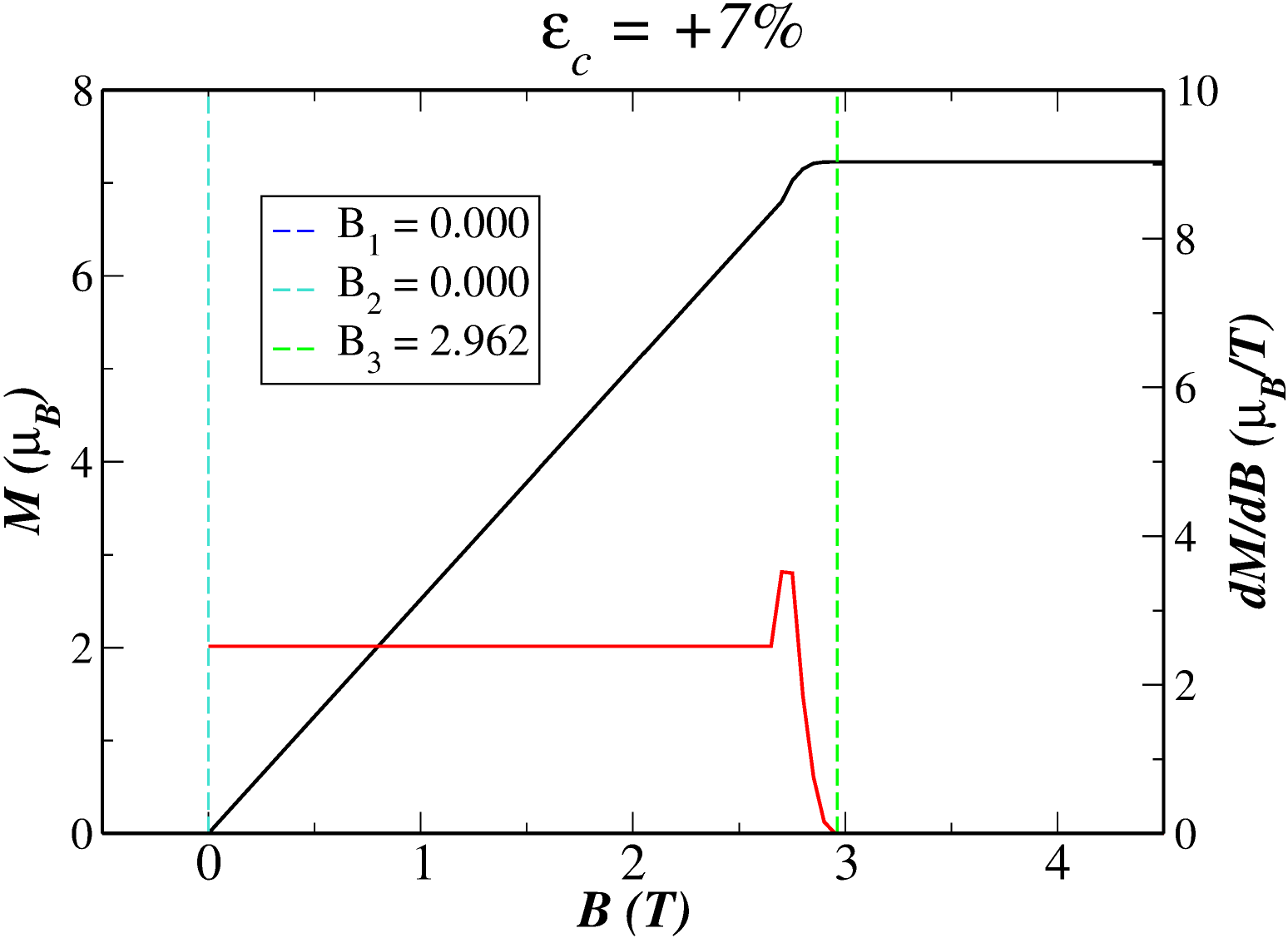}
\caption{System $\epsilon_c$ = $+7~\%$: The simulated $M(B)$ response and the corresponding $dM/dB$ data showing the position of peaks and determined critical fields $B_1$, $B_2$, and $B_3$.}
\label{derivative_p7}
\end{figure}
\FloatBarrier

\section{The methodology for Spin Structure Factor calculation}

The spin ordering or configurations in the 2D Gd layers in phases I, II, and III differ. The differences could be easily identified from the spin structure factor, the Fourier transform of the static spin-spin pair correlation function. This is defined as follows~\cite{hayami2021square}. 

\begin{equation}
 S^{\alpha}_s (\textbf{q}) = \frac{1}{N} \sum_{j,l}   S^{\alpha}_j S^{\alpha}_l e^{i\textbf{q} \cdot (\textbf{r}_j - \textbf{r}_l)}
 \label{eqn1}
\end{equation}

$S^{\alpha}_s$ stands for the $\alpha^\mathrm{th}$ component of the spin structure factor with $\alpha = x, y, z$. $N$ is the total number of spins in a single Gd layer. $S^{\alpha}_j$ is the $\alpha^\mathrm{th}$ component of the $j^\mathrm{th}$ spin located at the site $j$ with the position vector $\textbf{r}_j$. From here, we further define an in-plane (IP) and out-of-plane (OP) spin structure factor as defined in the following Eqns.~\ref{eqn2} and~\ref{eqn3}, respectively, to easily understand the spin modulations in different spin configurations. 

\begin{equation}
 S^{IP}_s (\textbf{q}) = S^{x}_s (\textbf{q}) + S^{y}_s (\textbf{q})
 \label{eqn2}
\end{equation}

\begin{equation}
 S^{OP}_s (\textbf{q}) = S^{z}_s (\textbf{q}) 
 \label{eqn3}
\end{equation}

\section{Spin configuration and Structure factor of magnetic phases}

From the calculated critical fields $B_1$, $B_2$, and $B_3$ as a function of strain, we have constructed a ($B ~\text{vs}~ \epsilon_c$) phase diagram consisting of different strain-driven and field-induced phases (see main text for details). In this section, we present the magnetic spin configurations and their characterization with the spin structure factors (Eqns.~\ref{eqn2} and ~\ref{eqn3}). The following figures (Figs.~\ref{PI_100} to \ref{can_fm}) show the same for Phases I, II, III, $\text{I}'$, $\text{II}'$, $\text{III}'$, and the canted FM state ($\text{FM}^c$). A $17\times17\times1$ square was cut from the actual 2D plane of the spins in each case for a better view of the structure.

\begin{figure}[ht]
\includegraphics[scale=0.55]{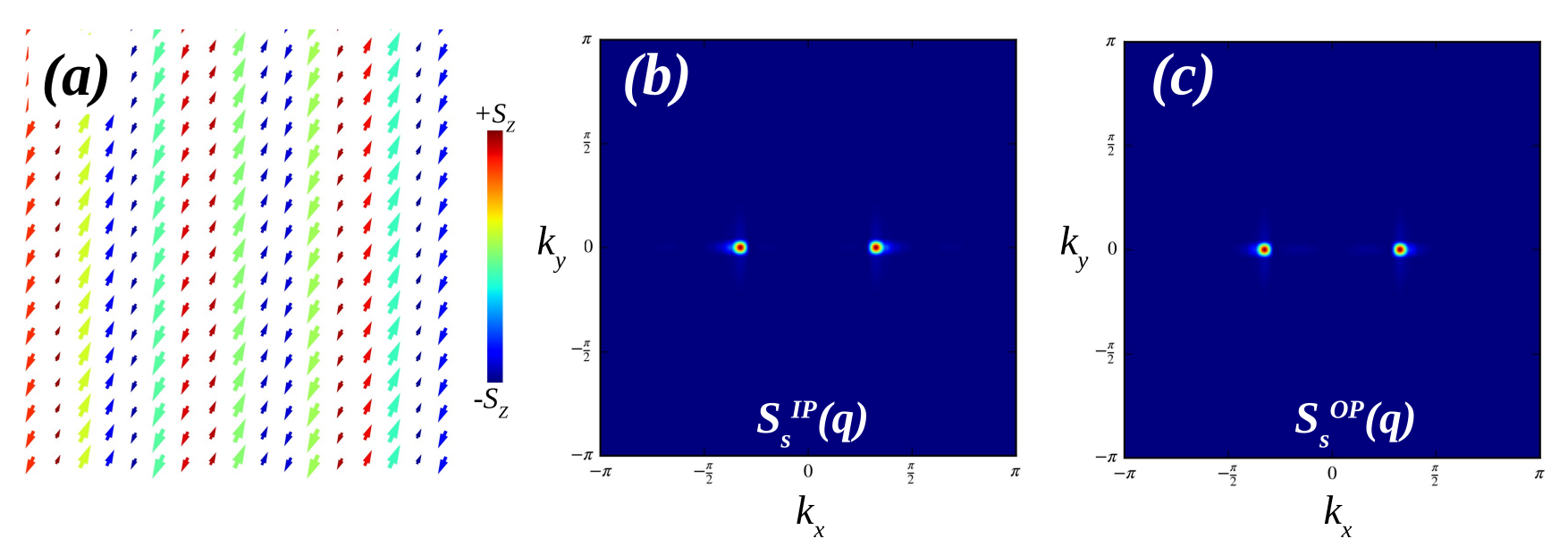}
\caption{Phase I: (a) Spin configuration within the 2D Gd layers. (b) The in-plane (IP) spin structure factor ($S^{IP}_s$) defined in Eqn.~\ref{eqn2} (c) The out-of-plane (OP) spin structure factor ($S^{OP}_s$) defined in Eqn.~\ref{eqn3}.}
\label{PI_100}
\end{figure}
\FloatBarrier

\begin{figure}[ht]
\includegraphics[scale=0.55]{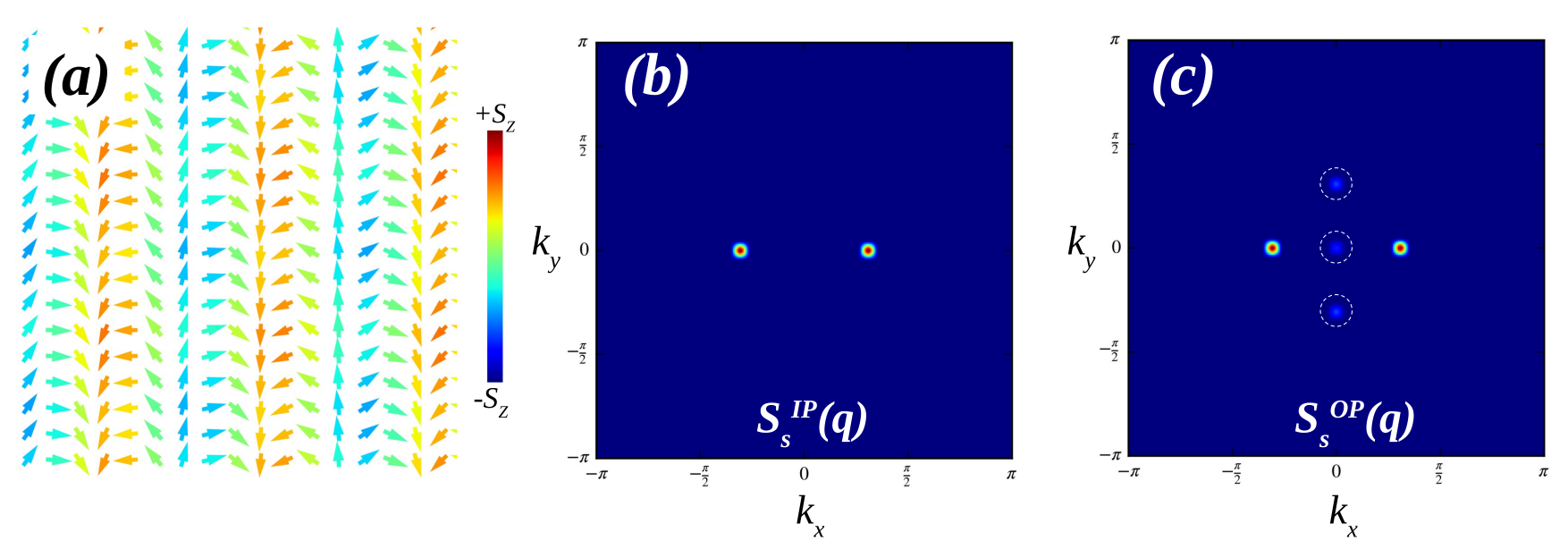}
\caption{Phase II: (a) Spin configuration within the 2D Gd layers. (b) The in-plane (IP) spin structure factor ($S^{IP}_s$) defined in Eqn.~\ref{eqn2} (c) The out-of-plane (OP) spin structure factor ($S^{OP}_s$) defined in Eqn.~\ref{eqn3}. The dotted circles indicate low-intensity peaks that are not very prominent.}
\label{PII_100}
\end{figure}
\FloatBarrier

\begin{figure}[ht]
\includegraphics[scale=0.55]{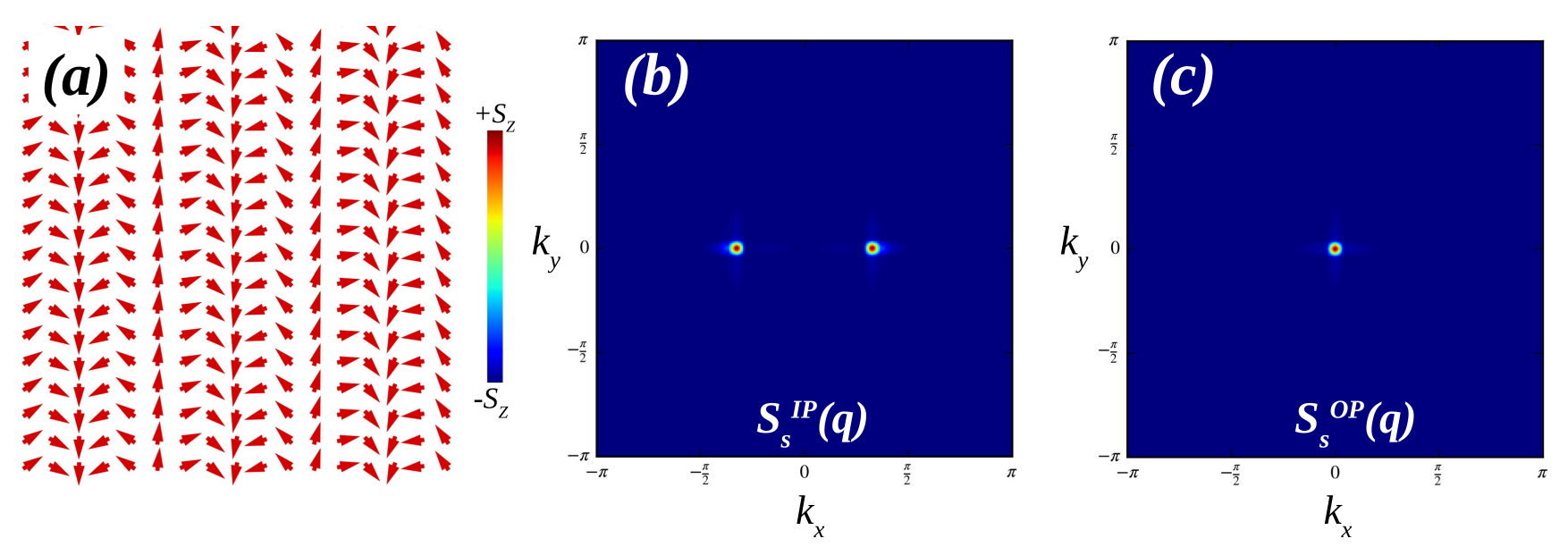}
\caption{Phase III: (a) Spin configuration within the 2D Gd layers. (b) The in-plane (IP) spin structure factor ($S^{IP}_s$) defined in Eqn.~\ref{eqn2} (c) The out-of-plane (OP) spin structure factor ($S^{OP}_s$) defined in Eqn.~\ref{eqn3}.}
\label{PIII_100}
\end{figure}
\FloatBarrier

\begin{figure}[ht]
\includegraphics[scale=0.55]{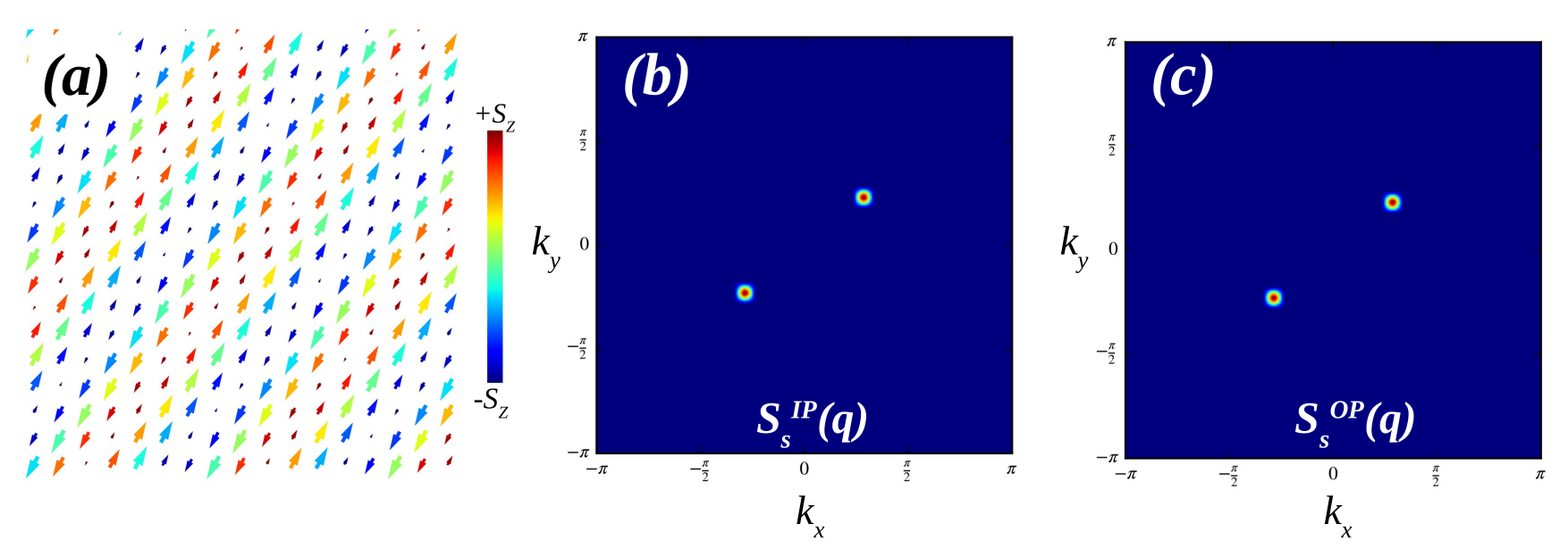}
\caption{Phase $\text{I}'$: (a) Spin configuration within the 2D Gd layers. (b) The in-plane (IP) spin structure factor ($S^{IP}_s$) defined in Eqn.~\ref{eqn2} (c) The out-of-plane (OP) spin structure factor ($S^{OP}_s$) defined in Eqn.~\ref{eqn3}.}
\label{PI_110}
\end{figure}
\FloatBarrier

\begin{figure}[ht]
\includegraphics[scale=0.55]{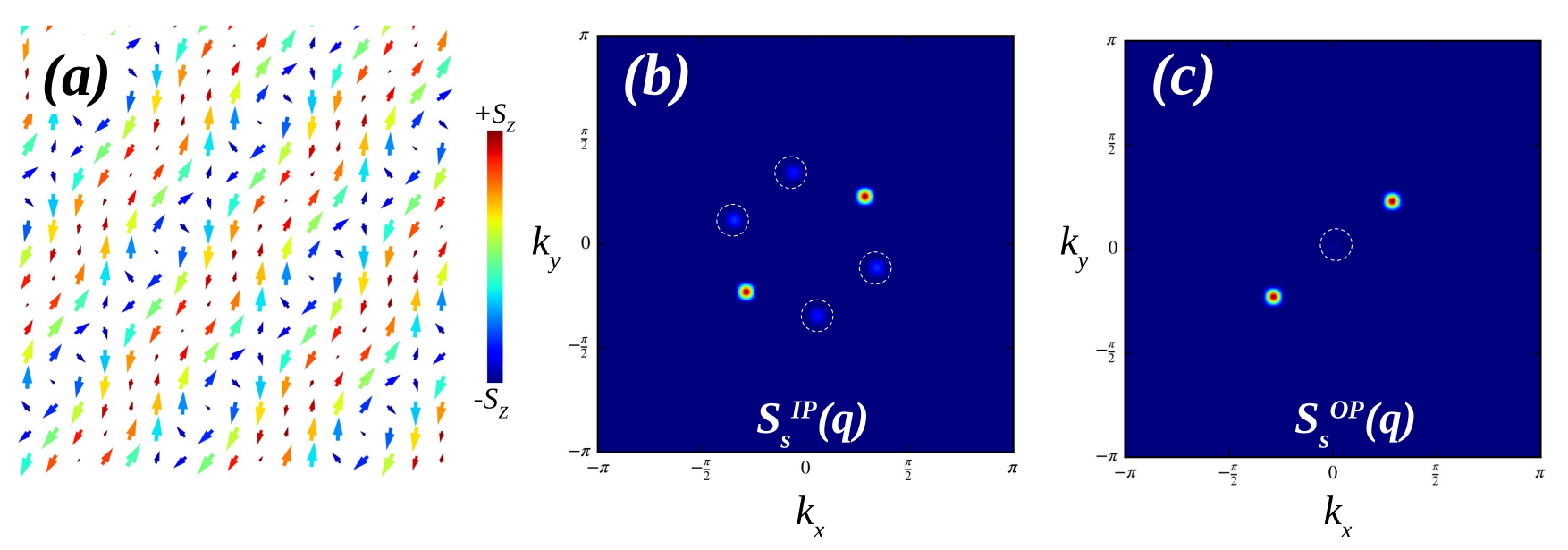}
\caption{Phase $\text{II}'$: (a) Spin configuration within the 2D Gd layers. (b) The in-plane (IP) spin structure factor ($S^{IP}_s$) defined in Eqn.~\ref{eqn2} (c) The out-of-plane (OP) spin structure factor ($S^{OP}_s$) defined in Eqn.~\ref{eqn3}. The dotted circles indicate low-intensity peaks that are not very prominent.}
\label{PII_110}
\end{figure}
\FloatBarrier

\begin{figure}[ht]
\includegraphics[scale=0.55]{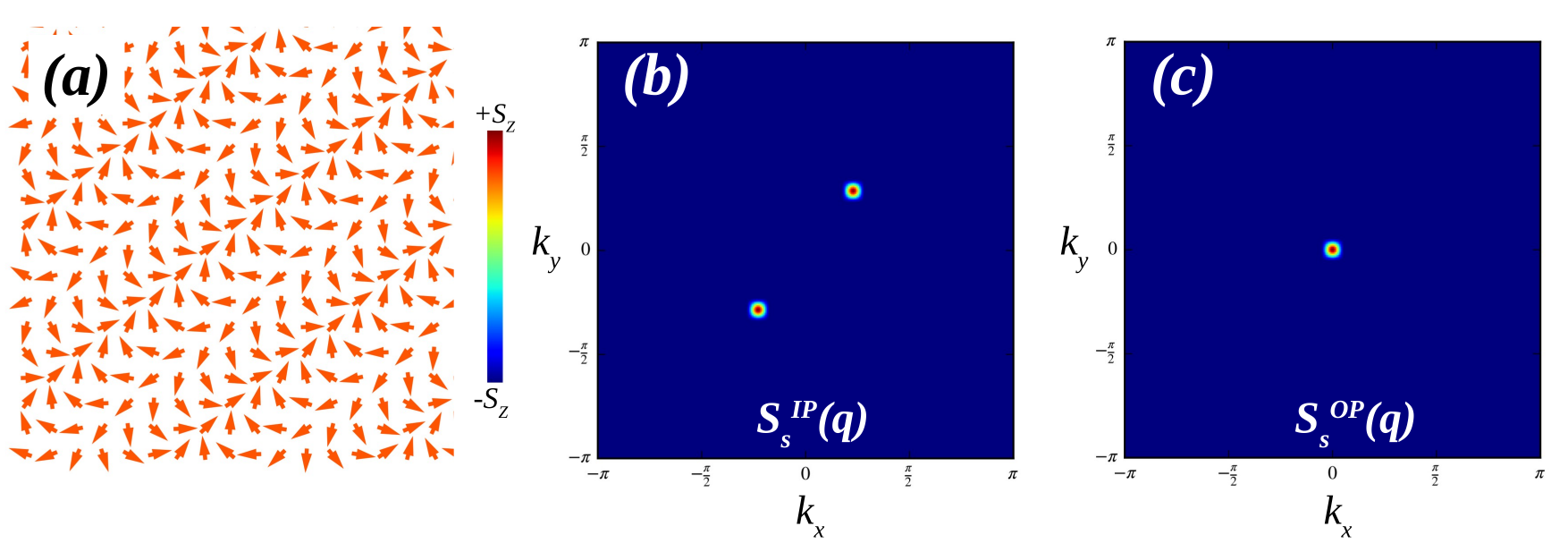}
\caption{Phase $\text{III}'$: (a) Spin configuration within the 2D Gd layers. (b) The in-plane (IP) spin structure factor ($S^{IP}_s$) defined in Eqn.~\ref{eqn2} (c) The out-of-plane (OP) spin structure factor ($S^{OP}_s$) defined in Eqn.~\ref{eqn3}.}
\label{PIII_110}
\end{figure}
\FloatBarrier

\begin{figure}[ht]
\includegraphics[scale=0.55]{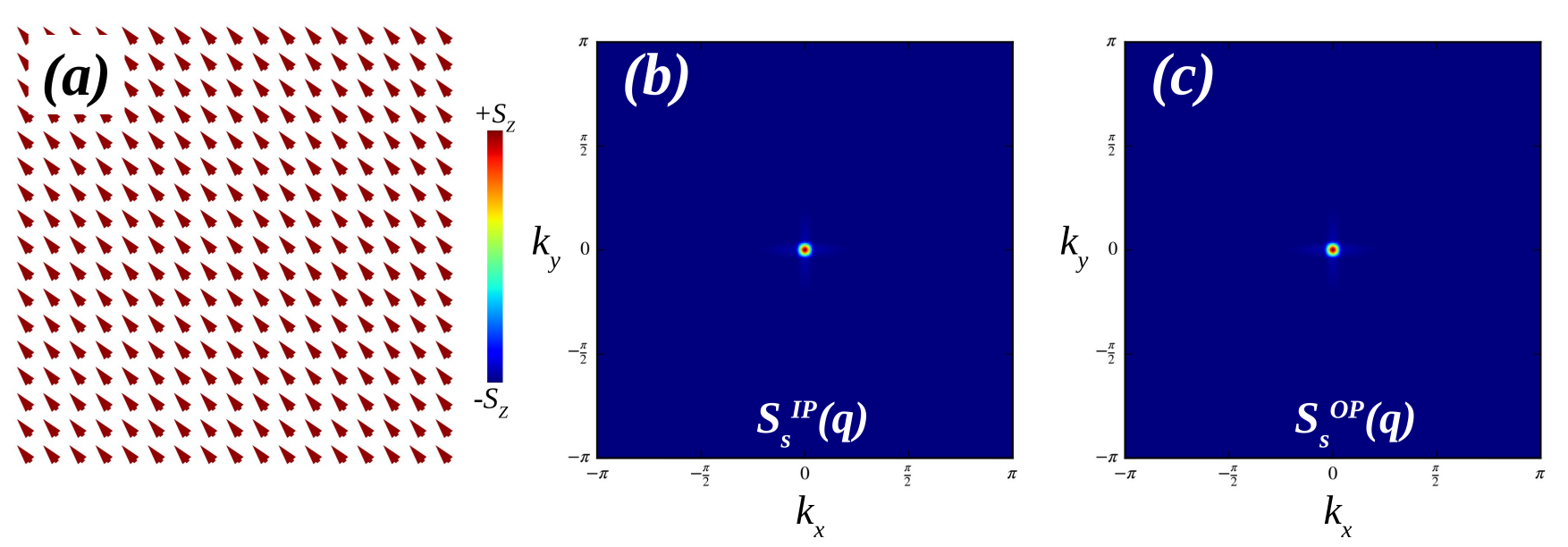}
\caption{Canted FM ($\text{FM}^c$) : (a) Spin configuration within the 2D Gd layers. (b) The in-plane (IP) spin structure factor ($S^{IP}_s$) defined in Eqn.~\ref{eqn2} (c) The out-of-plane (OP) spin structure factor ($S^{OP}_s$) defined in Eqn.~\ref{eqn3}.}
\label{can_fm}
\end{figure}
\FloatBarrier

\section{Effect of dipolar interactions on the spiral ground state}

\begin{figure}[ht]
\includegraphics[scale=0.55]{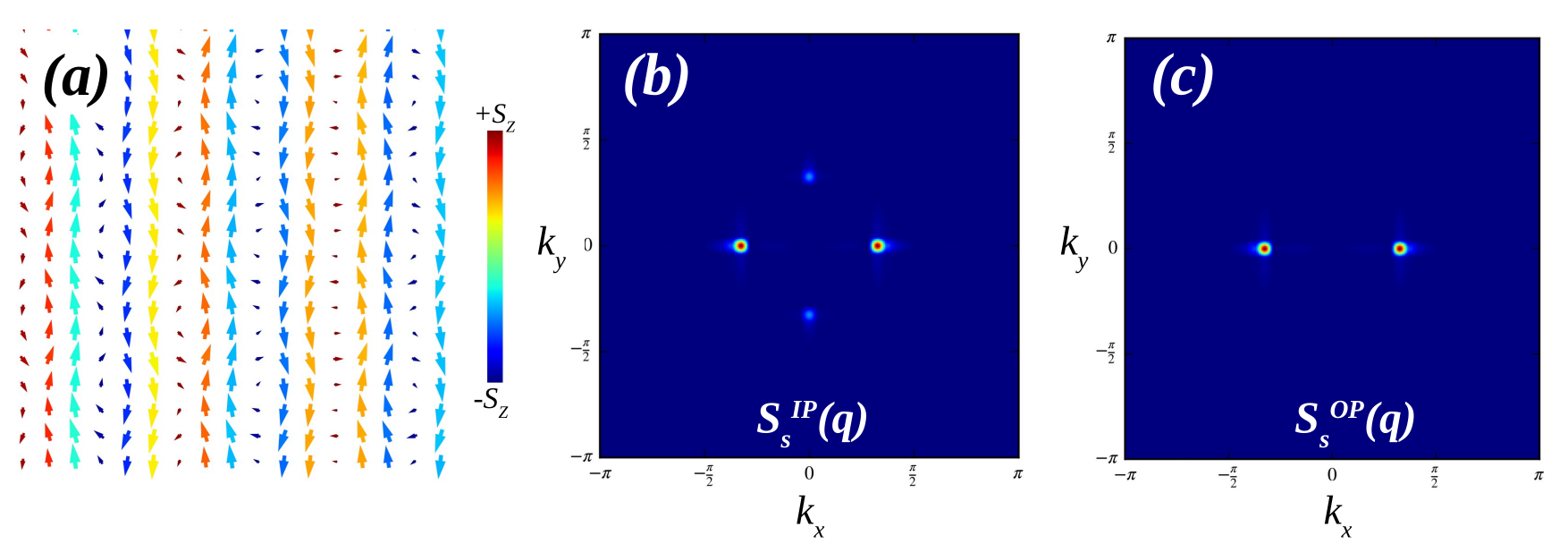}

\caption{A double-Q modulation of the in-plane spin components in Phase I due to dipolar interactions. (a) Spin configuration within the 2D Gd layers. (b) The in-plane (IP) spin structure factor ($S^{IP}_s$) defined in Eqn.~\ref{eqn2} (c) The out-of-plane (OP) spin structure factor ($S^{OP}_s$) defined in Eqn.~\ref{eqn3}.}
\label{dip-PI_100}
\end{figure}

\begin{figure}[ht]
\includegraphics[scale=0.55]{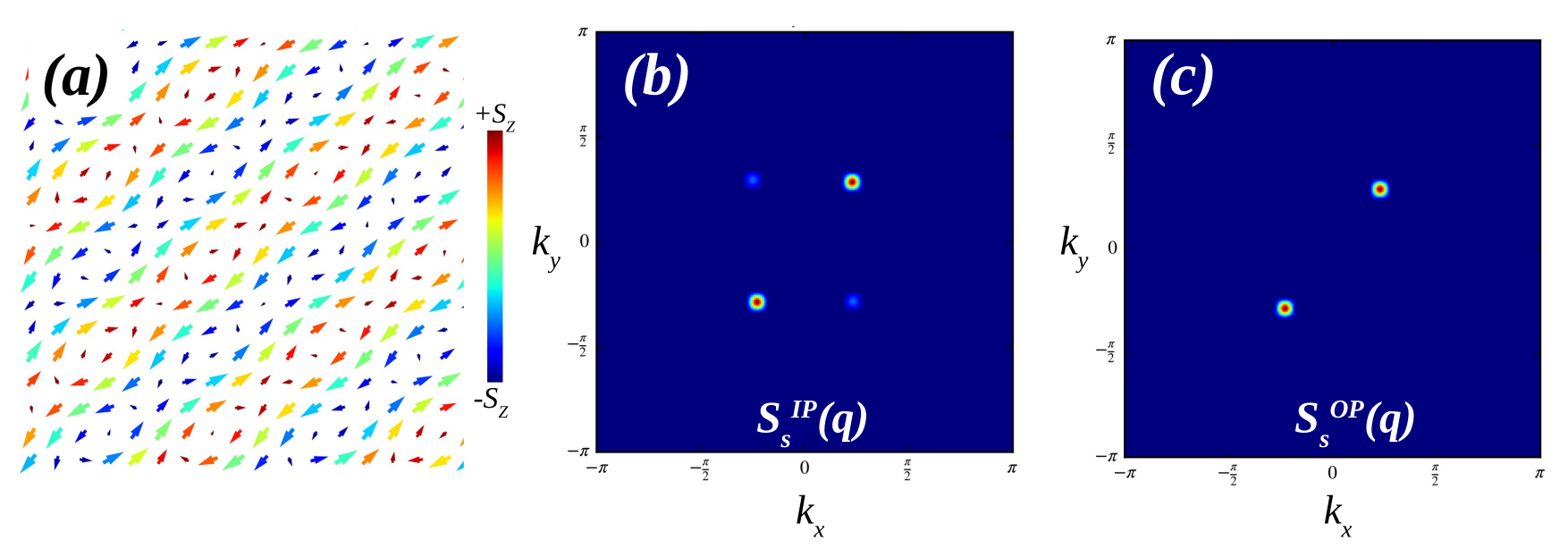}
\caption{A double-Q modulation of the in-plane spin components in Phase $\text{I}'$ due to dipolar interactions. (a) Spin configuration within the 2D Gd layers. (b) The in-plane (IP) spin structure factor ($S^{IP}_s$) defined in Eqn.~\ref{eqn2} (c) The out-of-plane (OP) spin structure factor ($S^{OP}_s$) defined in Eqn.~\ref{eqn3}.}
\label{dip-PI_110}
\end{figure}

\begin{figure}[ht]
\includegraphics[scale=0.55]{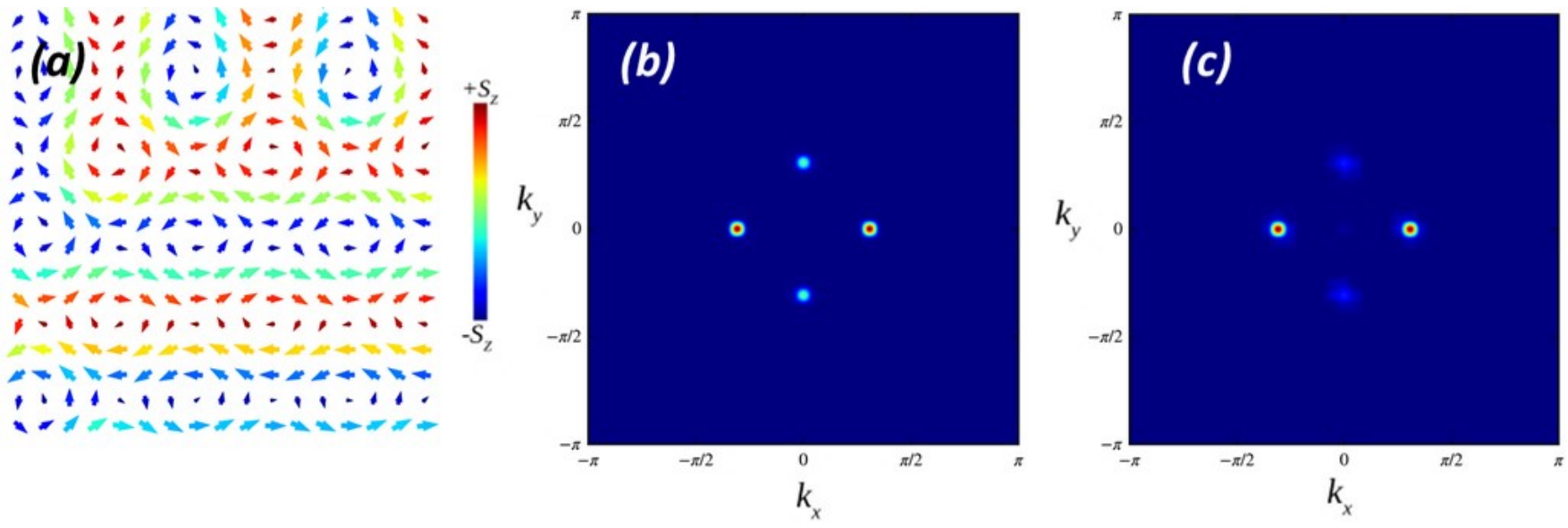}
\caption{A double-Q modulation of the in-plane spin components in Phase II due to dipolar interactions. (a) Spin configuration within the 2D Gd layers. (b) The in-plane (IP) spin structure factor ($S^{IP}_s$) defined in Eqn.~\ref{eqn2} (c) The out-of-plane (OP) spin structure factor ($S^{OP}_s$) defined in Eqn.~\ref{eqn3}.}
\label{dip-PII_100}
\end{figure}

\begin{figure}[ht]
\includegraphics[scale=0.55]{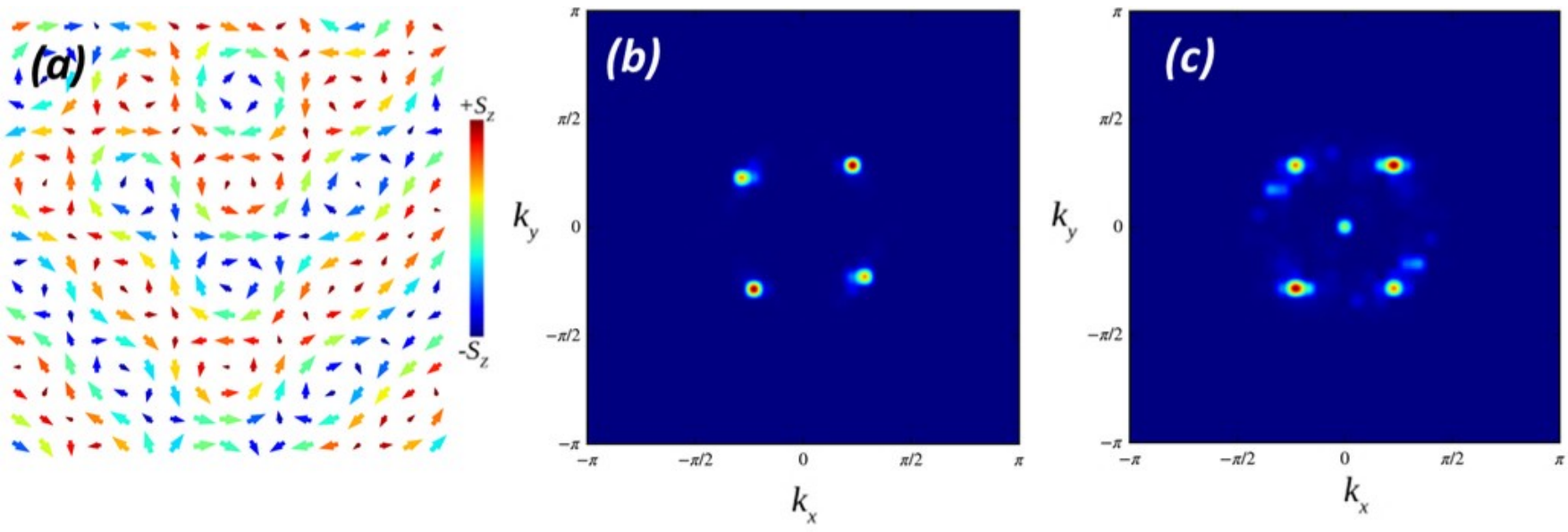}
\caption{A double-Q modulation of the in-plane spin components in Phase $\text{II}'$ due to dipolar interactions. (a) Spin configuration within the 2D Gd layers. (b) The in-plane (IP) spin structure factor ($S^{IP}_s$) defined in Eqn.~\ref{eqn2} (c) The out-of-plane (OP) spin structure factor ($S^{OP}_s$) defined in Eqn.~\ref{eqn3}.}
\label{dip-PII_110}
\end{figure}

\begin{figure}[ht]
\includegraphics[scale=0.55]{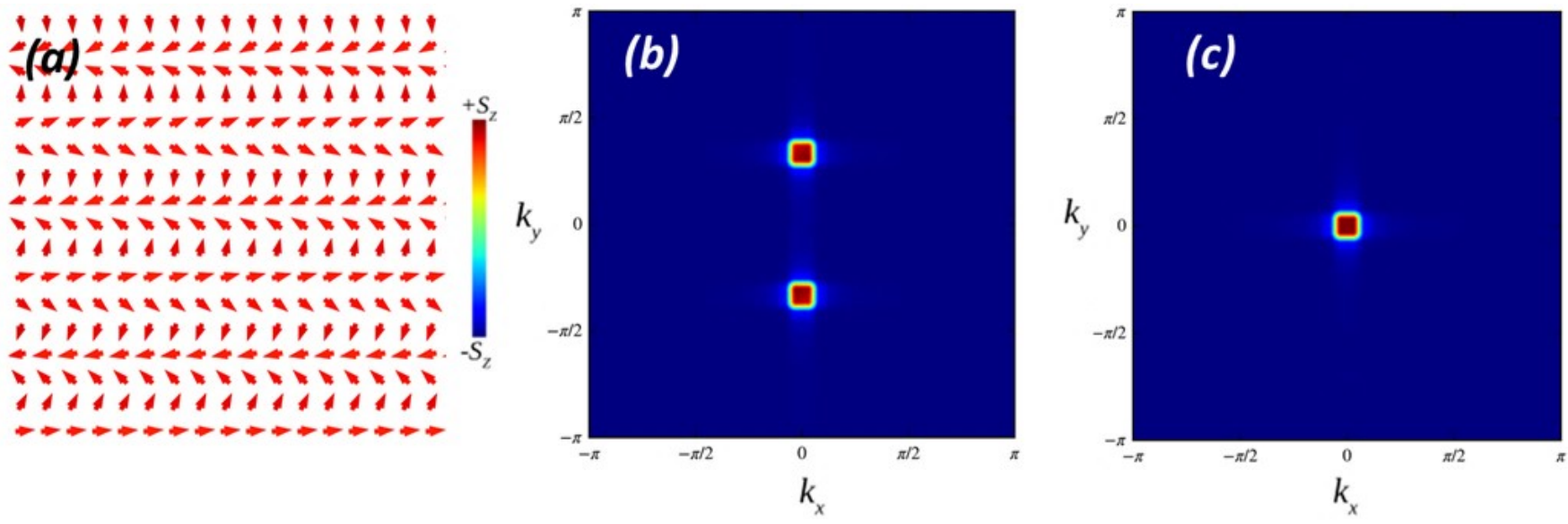}
\caption{Phase III with dipolar interactions. (a) Spin configuration within the 2D Gd layers. (b) The in-plane (IP) spin structure factor ($S^{IP}_s$) defined in Eqn.~\ref{eqn2} (c) The out-of-plane (OP) spin structure factor ($S^{OP}_s$) defined in Eqn.~\ref{eqn3}.}
\label{dip-PIII_100}
\end{figure}

\begin{figure}[ht]
\includegraphics[scale=0.55]{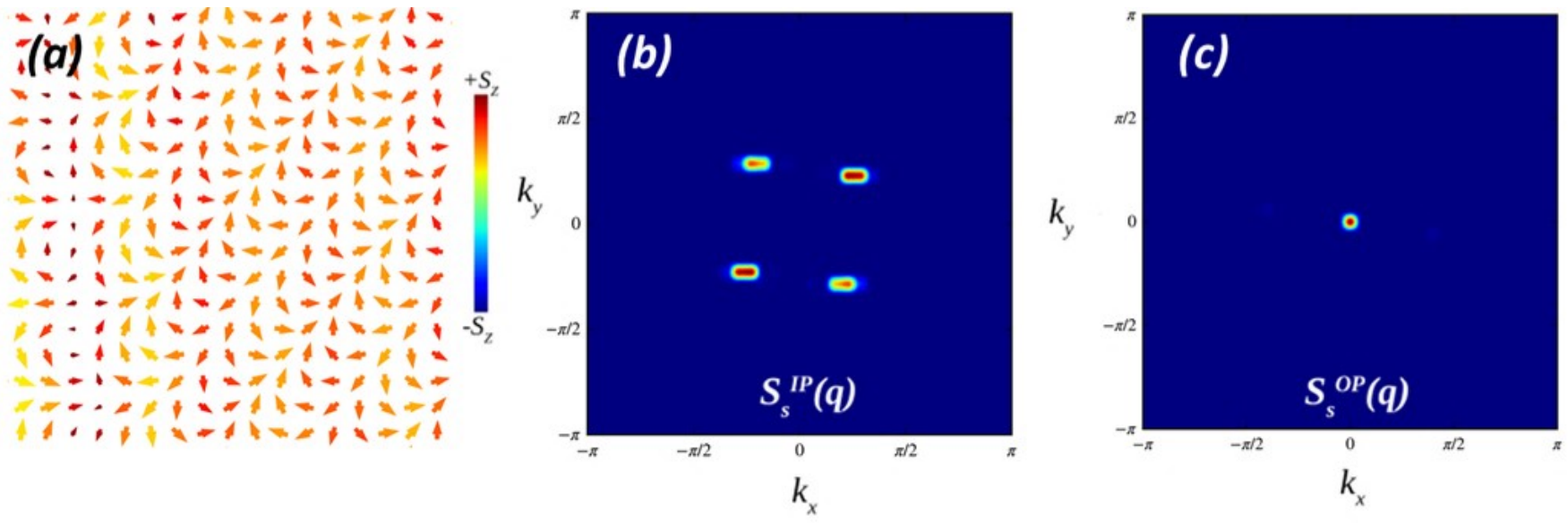}
\caption{A double-Q modulation of the in-plane spin components in Phase $\text{III}'$ due to dipolar interactions. (a) Spin configuration within the 2D Gd layers. (b) The in-plane (IP) spin structure factor ($S^{IP}_s$) defined in Eqn.~\ref{eqn2} (c) The out-of-plane (OP) spin structure factor ($S^{OP}_s$) defined in Eqn.~\ref{eqn3}.}
\label{dip-PIII_110}
\end{figure}

In our recent work on the magnetic properties of GdRu$_2$Si$_2$~\cite{SSarkar_PRB.112_2025}, we have already demonstrated the importance of dipolar interactions. It becomes essential for the correct zero-field magnetic ground state. However, as also claimed by Wood \textit{et al}~\cite{GDAWood_npjQM.10_2025}, in the ﬁeld-polarized phase, the effect of dipolar interactions is largely suppressed. As a result, the magnetic response of the system does not change qualitatively if dipolar interactions are included in our simulations~\cite{SSarkar_PRB.112_2025}. In this section, we present the impact of dipolar interactions on Phase I and $\text{I}'$ as shown in Fig.~\ref{dip-PI_100} and Fig.~\ref{dip-PI_110}, respectively. We can see additional modulation in the in-plane spin components induced by the dipolar effect, which is consistent with our previous findings~\cite{SSarkar_PRB.112_2025}. Next, we examine the impact on Phase II and $\text{II}'$ as shown in Fig.~\ref{dip-PII_100} and Fig.~\ref{dip-PII_110}, respectively. We can see additional modulation in the in-plane spin components induced by the dipolar effect. For Phase $\text{II}'$, we also observe additional modulation, and the in-plane and out-of-plane spin orientations interchange with dipolar interactions. Finally, we consider Phase III and $\text{III}'$ as shown in Fig.~\ref{dip-PIII_100} and Fig.~\ref{dip-PIII_110}, respectively. For Phase III, no major changes are observed as it is at high field, where field polarization does not give major changes with dipolar interactions. In contrast, for Phase $\text{III}'$, we again observe additional modulation in the in-plane structure factor plot. The methodology for the calculations of the dipolar interaction is provided in Ref.~\cite{SSarkar_PRB.112_2025}.

\FloatBarrier

\FloatBarrier

\FloatBarrier

\FloatBarrier

\FloatBarrier

\FloatBarrier

\medskip
\bibliography{main}